\documentclass{openjournal}

\usepackage{lipsum}

\usepackage{xcolor}
\usepackage{textgreek}
\usepackage[utf8]{inputenc}
\usepackage[english]{babel}

\usepackage{hyperref}
\hypersetup{
    unicode, 
    colorlinks=true,
    linkcolor=linkcolor,
    citecolor=linkcolor,
    filecolor=linkcolor,
    urlcolor=linkcolor,
}
\usepackage{color,colortbl}
\definecolor{linkcolor}{rgb}{0.0,0.3,0.5}
\usepackage{tensind}
\tensordelimiter{?}
\DeclareGraphicsExtensions{.bmp,.png,.jpg,.pdf}
\usepackage{verbatim}
\usepackage[normalem]{ulem}
\usepackage{orcidlink}
\usepackage{soul}
\usepackage{booktabs}
\usepackage{tabularx}
\usepackage{array}

\graphicspath{ {./figs/} }

\begin{document}
\title{The DSA/Chronoscope fast radio burst survey: forecasts and science overview}

\author{Liam Connor*\orcidlink{0000-0002-7587-6352}$^{1}$,
Kaitlyn Shin\orcidlink{0000-0002-6823-2073}$^{2}$,
Vikram Ravi\orcidlink{0000-0002-7252-5485}$^{2,3}$,
Stella Koch Ocker\orcidlink{0000-0002-4941-5333}$^{2,4}$,
Casey J. Law\orcidlink{0000-0002-4119-9963}$^{2,3}$,
Kritti Sharma\orcidlink{0000-0002-4477-3625}$^{2}$,
Samuel McCarty\orcidlink{0009-0008-5043-6220}$^{1}$,
Gregg Hallinan\orcidlink{0000-0002-7083-4049}$^{2,3}$,
Shami Chatterjee\orcidlink{0000-0002-2878-1502}$^{5}$,
James M. Cordes\orcidlink{0000-0002-4049-1882}$^{5}$,
Dean Howarth\orcidlink{0000-0002-9834-712X}$^{2}$,
Fabian Walter\orcidlink{0000-0003-4793-7880}$^{6}$,
Elisabeth Krause\orcidlink{0000-0001-8356-2014}$^{7,8}$,
Vishnu Balakrishnan\orcidlink{0000-0003-3244-2711}$^{1}$,
Alexa C. Gordon\orcidlink{0000-0002-5025-4645}$^{9}$,
Calvin Leung\orcidlink{0000-0002-4209-7408}$^{10,11}$,
Shion Andrew\orcidlink{0000-0002-3980-815X}$^{12,13}$\vspace{0.5em}
}

\affiliation{
  $^{1}$Center for Astrophysics $|$ Harvard \& Smithsonian, 60 Garden Street, Cambridge, MA 02138, USA\\
  $^{2}$Cahill Center for Astronomy and Astrophysics, MC 249-17, California Institute of Technology, Pasadena, CA 91125, USA\\
  $^{3}$Owens Valley Radio Observatory, California Institute of Technology, Big Pine, CA 93513, USA\\
  $^{4}$The Observatories of the Carnegie Institution for Science, 813 Santa Barbara Street, Pasadena, CA 91101, USA\\
  $^{5}$Cornell Center for Astrophysics and Planetary Science and Department of Astronomy, Cornell University, Ithaca, NY 14853, USA\\
  $^{6}$Max-Planck-Institut f\"ur Astronomie, K\"onigstuhl 17, 69117 Heidelberg, Germany\\
  $^{7}$Department of Physics, University of Arizona, Tucson, AZ 85721, USA\\
  $^{8}$Steward Observatory, University of Arizona, 933 North Cherry Avenue, Tucson, AZ 85721, USA\\
  $^{9}${Center for Interdisciplinary Exploration and Research in Astrophysics (CIERA) and Department of Physics and Astronomy, Northwestern University, Evanston, IL 60208, USA}\\
  $^{10}$Miller Institute for Basic Research, University of California, Berkeley, CA 94720, USA\\
  $^{11}$Department of Astronomy, University of California, Berkeley, CA 94720, USA\\
  $^{12}$MIT Kavli Institute for Astrophysics and Space Research, Massachusetts Institute of Technology, 77 Massachusetts Ave, Cambridge, MA 02139, USA\\
  $^{13}$Department of Physics, Massachusetts Institute of Technology, 77 Massachusetts Ave, Cambridge, MA 02139, USA\\
  }
\thanks{Corresponding author: Liam Connor:\\ \href{mailto:liam.connor@cfa.harvard.edu}{liam.connor@cfa.harvard.edu}}


















\begin{abstract}
    Fast radio bursts (FRBs) are bright extragalactic transients with several mysteries surrounding their origins. Large FRB samples enable accurate measurements of the cosmic matter distribution, in particular on scales $\lesssim 10$\,Mpc. These measurements will impact cosmological inference and our understanding of astrophysical feedback, from the circumgalactic medium to cluster scales. Here we forecast the expected yields, redshifts, and host galaxies of FRBs as observed by the Deep Synoptic Array (DSA), and describe the key science cases enabled by the large FRB sample. The DSA will be an interferometer consisting of 1650$\times$6.15\,m antennas, operating between 0.7--2\,GHz, to be located in Nevada, USA. The Chronoscope backend on the DSA, hereafter DSA/Chronoscope, is designed to search for FRBs across the field of view in real time, enabling the storage of full-polarization voltage data. Extrapolating from existing FRB surveys, we expect roughly $10^4$ FRB detections per year in each of three search sub-bands. Combining across sub-bands, the survey could produce $\sim$\,10$^5$ FRBs over the nominal 5-year DSA survey, assuming Euclidean source counts and a baseline compute backend that can search $6\times10^6$ beams at 1\,ms sampling.
    The well-characterized DSA synthesized beam and deep simultaneous reference images will enable localization precisions of $\lesssim$\,250\,milliarcseconds. Key science cases include the use of FRB propagation effects in probing cosmic baryons, and studies of the FRB phenomenon using FRB host galaxies and their local environment, as well as multiwavelength counterparts. 
\end{abstract}


\maketitle

\newcommand{\dmunits}{\,pc\,cm$^{-3}$}

\section{Introduction}
\label{sec:intro}

Fast radio bursts (FRBs) are typically sub-second, luminous ($\sim10^{35}-10^{43}$\,erg) phenomena observed across the Universe \citep{lorimer2007, petroffreview}. A Galactic event similar to extragalactic FRBs was associated with a magnetar \citep{Bochenek2020STARE2, Andersen2020SGR1935CHIME}, and a plurality of extragalactic FRBs originate from galaxies (and, where data are available, environments) consistent with Galactic magnetars. However, outlier events in, for example, a globular cluster \citep{KirstenM81} and embedded within persistent radio sources in dwarf galaxies \citep{chatterjee_direct_2017, Niu2022, moroianu_milliarcsecond_2025} indicate a diversity of so far unobserved source formation channels. Further, FRB phenomenology and energetics require extreme electrodynamical processes \citep{kumar}. Identifying the sources and emission mechanisms of FRBs will rely on a battery of breakthroughs: the discovery of very nearby sources that have the potential for multiwavelength characterization of the engine, transient emission, and immediate environments; the conclusive identification of source classes and evolution in the population; and the discovery of sources that redefine the limits of the population in energy release, spectro-temporal burst properties, and location. 

Data on FRBs carry the imprints of plasma and compact structures along their sightlines \citep{cordesreview}. Propagation effects in FRB data probe the column densities, inhomogeneity, and magnetization of the tenuous plasma around and in between galaxies that comprises the overwhelming majority of cosmic baryons \citep{connor2025}. FRBs are now established as a forefront probe of the baryon contents of galaxy halos and the cosmic web \citep{macquart2020, connor2025, sharma2026backlightingcosmicwebfast}, and also enable precise measurements of baryonic effects on the matter power spectrum on $\sim$Mpc scales \citep{hussaini25, wang25,sharma2026signaturessuppressedmatterclustering}. In combination with Stage IV cosmology surveys, FRBs will transform our understanding of how astrophysical feedback couples to baryons, and will uniquely enable cosmological inference on small scales \citep{sharma2026signaturessuppressedmatterclustering}. Dramatic improvements are anticipated in measurements of the sum of neutrino masses, the strength of matter clustering, and dark energy and its evolution \citep{reischke26a}. FRB propagation also enables gravitational-lensing studies via time delays of extragalactic milli- and micro-lenses, with important implications for the nature of dark matter \citep{connorravilens}.

To fully capitalize on the scientific potential of FRB observations requires an unprecedented sample of events. The FRB sample obtained by the Canadian Hydrogen Intensity Mapping Experiment (CHIME) telescope contains well over 5000 events localized at the $\gtrsim$\,arcminute level \citep{frbcollaboration2026secondchimefrbcatalogfast}, and interferometers like the 110-antenna Deep Synoptic Array \citep[DSA-110;][]{law2024, sharma_preferential_2024, connor2025}, the Australian SKA Pathfinder \citep[ASKAP;][]{shannon2025craft}, CHIME and its outrigger antennas \citep{chimevlbi}, and MeerKAT \citep{Jankowski_2023} have combined to discover nearly 200 events localized with $\sim$arcsecond precision. These data have led to the discovery of unusual individual events and objects, statistical population modeling of sources and their propagation effects, and via the association of dispersion measures with tracers of baryons in halos and the intergalactic medium. The results to date highlight the need for orders of magnitude larger FRB samples to address the problems of FRB sources and emission, and to fully realize the potential of FRBs as probes.

This paper describes, forecasts, and motivates the FRB survey that the Deep Synoptic Array \citep[DSA;][]{hallinan2019} will undertake. Fundamentally, telescopes that seek to maximize the detection rate of FRBs with sufficient localization precision to enable their full exploitation need to excel along three axes: sensitivity, angular resolution, and field of view (FoV). Prioritizing FoV with angular resolution offers the opportunity to discover and characterize the nearest, rarest FRB sources, addressing critical questions on FRB engines and emission mechanisms. This technical niche is currently the domain of CHIME/FRB Outriggers program \citep{chimevlbi}, and will ultimately be filled by the next generation of dense aperture arrays being developed for FRB searches, including the Coherent All-Sky Monitor \citep[CASM;][]{casm}, the Bustling Universe Radio Survey Telescope in Taiwan \citep[BURSTT;][]{BURSTT}, the Canadian-Chilean array for radio transient studies \citep[CHARTS;][]{charts}, and similar projects elsewhere. The DSA, on the other hand, will have comparable sensitivity to the Five-hundred-meter Aperture Spherical Telescope \citep[FAST;][]{fast}, with sub-arcsecond FRB localization accuracy, and sufficient FoV to provide the requisite few orders of magnitude increase in FRB detection rate. The DSA events will be ideally suited to delivering on the full breadth of possibilities with FRBs as probes, and to characterizing and classifying FRB sources over cosmic time. Upcoming projects such as the 512 6.1\,m array CHORD \citep{chord} and SKA \citep{caleb2026probingbaryondistributionfast} are also expected to detect large numbers of FRBs in the 2030s.

The DSA, formerly known as the DSA-2000, is a radio observatory that will be located in Nevada, U.S., with operations expected to start in 2029. The array will consist of 1,650 6.15-m fully steerable dishes located within a $\sim 20 \times 16$\,km region. Each dish will be equipped with a dual-polarization 0.7--2\,GHz ambient-temperature receiver, with typical band-averaged system temperatures at zenith around 21\,K and an aperture efficiency of 0.67 \citep{flygareDSATsys}. A nominal full-array system-equivalent flux density (SEFD) at zenith of 1.8\,Jy is expected. Analog signals from each antenna will be transmitted over fiber to a central digitizer/channelizer, which will in turn stream data to parallel digital backends. The Radio Camera digital backend replaces the traditional correlator backend of a radio telescope. It integrates an F-X correlator with streaming calibration, data flagging, bright source removal (peeling) and gridding/imaging, and will be implemented on RFSoC FPGA system-on-chips and data center GPUs with high-bandwidth interconnect. 
The Radio Camera is an imaging backend, delivering spectral image cubes spanning the full bandwidth that are intended to reach the theoretical imaging sensitivity, fidelity, and dynamic range, serving continuum, spectral-line, polarization, and ``slow'' ($\gtrsim10$\,s) time-domain science cases. It also computes time-varying direction-dependent calibration solutions during every observation, correcting for ionospheric phase-screen variations across the array. The Pulsar Timing backend processes four tied-array beams, also spanning the full bandwidth, applying coherent dedispersion and folding routines for high-precision timing cases such as low-frequency gravitational-wave detection. The channelized data streamed into the Radio Camera GPUs are also sent to a second GPU-based backend called the Chronoscope\footnote{A conceptual science fiction device allowing one to observe another point in time, such as in Isaac Asimov's short story \textit{The Dead Past} \citep{chronoscope_asimov}.}. The Chronoscope/FRB system performs a full-field, full-sensitivity search for FRBs and other $\lesssim10$\,s single pulses in $\sim50\%$ of the DSA bandwidth, as described below, together with an all-sky pulsar survey with the Chronoscope/PSR system.  


We begin in Section~\ref{sec:survey_overview} with a summary of the Chronoscope/FRB (CHR/FRB) specifications, design, and data products. Full descriptions of the CHR/FRB system and the DSA, and the enabling technologies, will be the subjects of future publications. The summary we present is based on a reference design, pending final hardware and software implementations, and the final computing capabilities at the disposal of the Chronoscope. We emphasize that more computing hardware capabilities enable more science, and we are in an era where the compute capacity (operations, memory, memory bandwidth) per unit cost is increasing sharply, although fluctuating dramatically. In Section~\ref{sec:forecasts}, we present a forecast for the extragalactic FRB population detected by CHR/FRB, including detection rates, the source redshift distribution, and prospects for accurate localization and host-galaxy characterization. In Section~\ref{sec:frb_science}, we present an overview of the science cases enabled by CHR/FRB data. We conclude in Section~\ref{sec:concl}. Throughout, we assume the Planck 2018 cosmological parameters \citep{Planck2020}. 

\begin{figure}[htbp]
    \centering
    \includegraphics[width=0.87\textwidth]{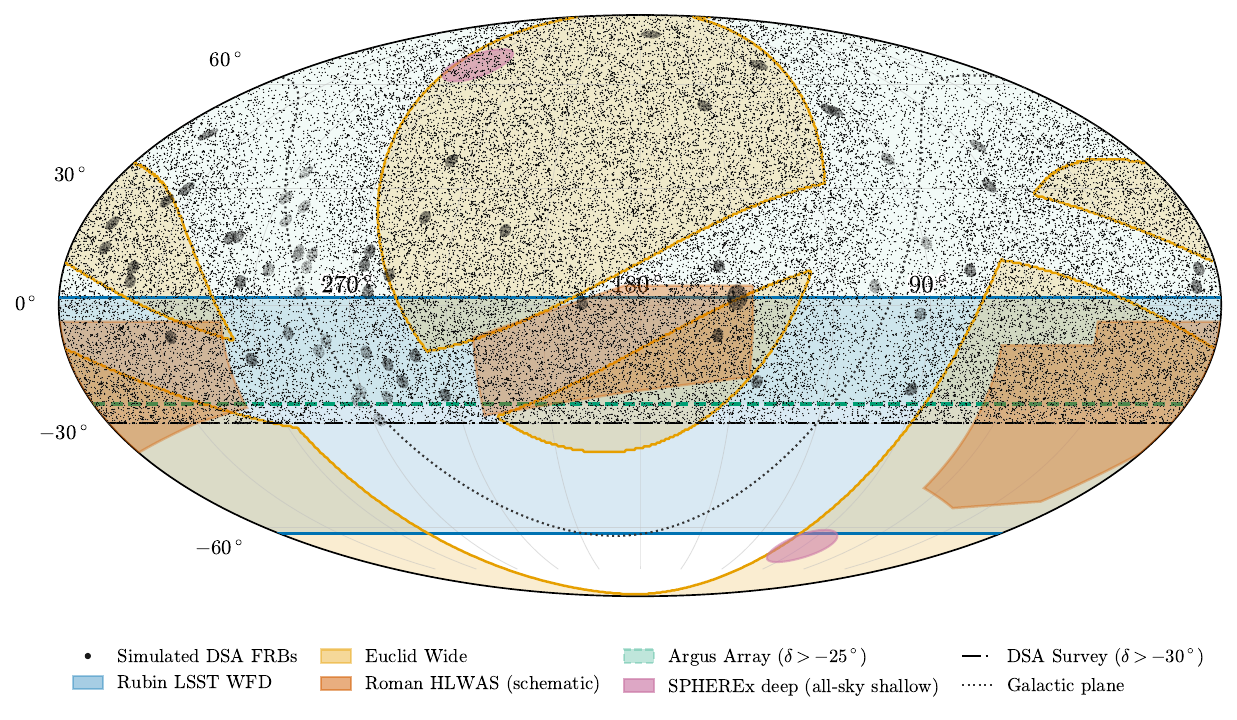}
    \caption{A simulated all-sky map of 50,000 DSA Chronoscope FRBs (black points) including the putative survey footprints of photometric O/IR surveys for host galaxy follow up. Overdense FRB regions correspond to existing NANOGrav pulsar timing and possible deep drilling fields, which will receive more exposure than typical fields. The number of pulsar timing fields and corresponding FRB overdensity shown is for visualization purposes and is expected to change in the final survey design. We assume decreased FRB detection rates at low Galactic latitudes. The Argus Array footprint is well matched to the DSA footprint, shown in the light green shaded region at Dec\,$\gtrsim-20^\circ$ \citep{Law_2022}. SPHEREx and the proposed NANCY survey with Roman cover $4\pi$\,sr and are not shown explicitly.}
    \label{fig:allsky}
\end{figure}
\vspace{5mm}

\vspace{1cm}
\section{Chronoscope/FRB survey summary}
\label{sec:survey_overview}

The CHR/FRB system will be used to search for and characterize FRBs across the $\sim2.5^{\circ}$ DSA primary beam, using a large number of regularly spaced beams ($\sim6 \times 10^6$). With the approximately  16$\times$20\,km DSA configuration, the synthesized beamwidth is $\lesssim$\,3\arcsec~in the middle of the 0.7--2\,GHz DSA band for zenith angles $<40^{\circ}$. CHR/FRB observations are commensal with all other DSA operations. Further detailed information about the CHR/FRB system will be presented in a separate paper. In the remainder of this section, we summarize the essentials of the CHR/FRB system and survey that are necessary for forecasting the scientific return. 

In Table~\ref{tab:frb_specs}, we provide some fiducial specifications for the FRB search enabled by the currently planned baseline compute. We emphasize that exact parameters are flexible, and will be further optimized during commissioning to maximize scientific return. The range of DMs given is conservative and encompasses all three CHR/FRB bands. The actual range of searched DMs searched for each band will depend on optimizations driven by the available memory and compute capabilities of the final hardware platform.

\begingroup
\renewcommand{\colhead}[1]{\multicolumn{1}{l}{#1}}
\begin{deluxetable}{ll}[htpb]
\tablewidth{0pt}
\tablecaption{Baseline specifications of the CHR/FRB search, which is done on intensity data. \label{tab:frb_specs}}
\tablehead{
    \colhead{\textbf{Parameter}} & \colhead{\textbf{Value}}
}
\startdata
     System equivalent flux density (SEFD) & 1.8\,Jy \\
     CHR/FRB Band~1 & 880--980\,MHz \\
     CHR/FRB Band~2 & 1310--1510\,MHz \\
     CHR/FRB Band~3 & 1650--1950\,MHz \\
     Number of beams & $6 \times 10^6$ \\
     Field of view &  $8.6 \,\nu_{\rm GHz}^{-2}\,\rm deg^2$ \\
     Temporal resolution & 1\,ms \\
     Spectral resolution & $\sim122$\,kHz \\
     Polarizations & I \\
     Number of bits in intensities & 4 \\
     Maximum searched dispersion measure (DM) & $\sim800-5000$\,pc\,cm$^{-3}$
\enddata
\end{deluxetable}
\endgroup

The Chronoscope ingests a copy of a $\sim122$\,kHz channelized voltage data stream for all DSA antennas, spanning the full band. After calibrations are applied and radio-frequency interference (RFI) signals are excised, the data will be beamformed using a post-correlation approach \citep{pcbeamforming}, generating intensity data integrated over 1\,ms. These data will be de-dispersed using a custom implementation of the Fourier-domain dedispersion algorithm accelerated using tensor cores \citep{bassa_fdd}, with up to 2000 optimally spaced \citep{keane_dmtol} DM trials, and searched in real-time using an AI/ML-based pipeline.  

In order to mitigate the challenge of processing large data volumes at the desired temporal and spectral resolution, rather than search the full DSA frequency range of 0.7--2\,GHz, the Chronoscope will search three discrete frequency bands: 880--980\,MHz, 1310--1510\,MHz, and 1650--1950\,MHz.
The frequency bands were chosen to be the least affected by RFI, identified via early RFI surveys at the planned DSA site. Our forecasted expectations are that sufficiently many FRBs will be broadband enough such that projected detection rates will not be too affected (further details in Section~\ref{sec:forecasts}).
We also know that while most FRBs are broadband, a non-trivial number of FRBs --- particularly repeaters --- exhibit narrow-bandedness \citep{Pleunis_2021b}.
Thus, the discrete frequency bands were also spaced with the goal of being more sensitive to narrow-banded FRBs across a broad range of frequencies.

The CHR/FRB survey will distribute public FRB alerts for high-probability FRB candidates, for example via the General Coordinates Network\footnote{\url{https://gcn.nasa.gov/}}.
Fast public alerts will include the maximum-S/N beam position, DM, arrival time, information on whether the candidate is likely a repeater, and a metric for astrophysical/FRB probability.
At a slower latency, further refined localizations from voltage data will be broadcast (Section~\ref{subsec:localization_precision}). Astrometric precision is assumed to be roughly the PSF scale divided by 2$\times$S/N of the FRB, down to a design threshold astrometric precision of 50\,mas.
Long-term data products will also be archived for all high-probability FRB sources that meet the criteria of triggering an alert, as well as interesting Galactic pulses, and astrophysical but ambiguously extragalactic bursts.
These data products include filterbank data at millisecond resolution which will be preserved for the detection beam and a subset of adjacent beams.
Voltage data across the full 0.7--2\,GHz band will also be saved to disk.
At DM=3000\dmunits, the delay across the full band 
is $\sim$\,22\,s, totaling 180\,TB of data --- too large to routinely write to disk.
For this reason, the pulse will be cut out in time and frequency, saving $\approx$\,4\,s around the center of the burst in each frequency channel.
Preserved voltage data will enable several science cases, including improved localization, searches for strong gravitational lensing and plasma lensing, high time- and frequency-resolution studies, and even VLBI in the future.

\begin{figure}[htbp]
    \centering
    \includegraphics[width=0.7\textwidth]{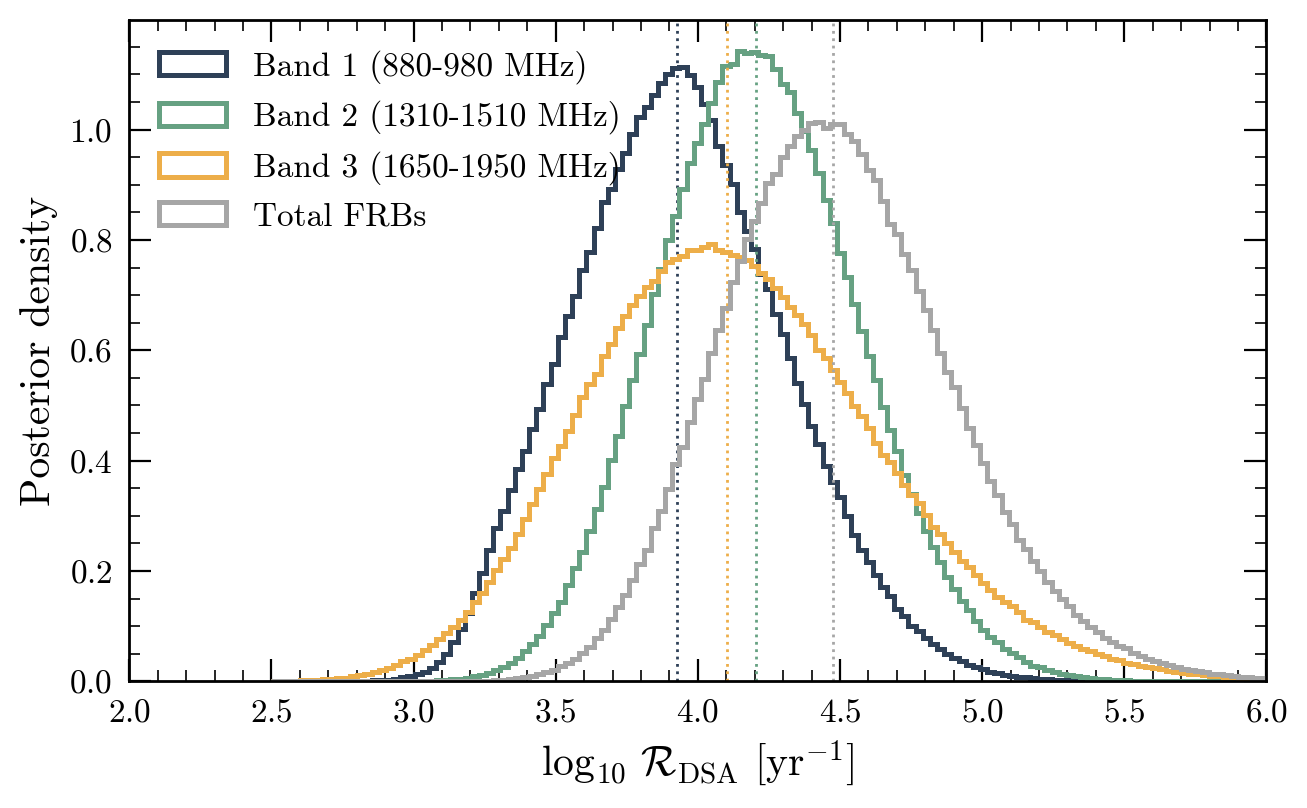}
    \caption{Probability distributions over the expected DSA/Chronoscope FRB detection rate in its three search sub-bands. Rate estimates are extrapolated from existing surveys such as CHIME/FRB (S1) and Murriyang HTRU (S2 \& S3). Uncertainties in telescope parameters, rate spectral index, and source counts explain the widths of these distributions. For each sub-band, the expectation is roughly $10^4$ FRBs 
    per year. The total rate after combining across sub-bands is shown in grey.}
    \label{fig:ratesMC}
\end{figure}

\vspace{1cm}
\section{Forecasts}
\label{sec:forecasts}
We forecast the DSA Chronoscope FRB survey's 
detection rates, localization precision, redshift distribution, and host galaxy magnitudes. We assume the baseline GPU-based compute backend that 
can coherently tile the full DSA FoV and search several million beams at millisecond time resolution. We note that compute hardware falling below this baseline can still accommodate the full-FoV search by forming fewer beams with $uv$ plane tapering, reducing time/frequency resolution, or altering the DM search plan. We comment on the sensitivity reduction under such scenarios in the following section. Equivalently, compute hardware falling above the baseline will directly enable higher sensitivity searches with more complete coverage of the search parameter space.

\subsection{Detection rates}
We forecast the CHR/FRB detection rate for each 
sub-band, after which we estimate the total number of unique 
bursts and sightlines discovered per year. We elect to scale from 
empirical detection rates at other surveys with similar frequency 
bands. This is more reliable than forward modeling from 
poorly-constrained volumetric FRB rates, luminosity functions, and assumed redshift 
distributions, because other surveys suffer from a similar set of 
selection effects and real-world performance issues. 
Throughout, we consider the FoV of DSA to 
be the area of an individual antenna's primary beam within its 
full-width half-max (FWHM), 

\begin{equation}
    \Omega_{\rm DSA} \approx 8.6 \,\nu_{\rm GHz}^{-2}\,\rm deg^2
\end{equation}

\noindent Scaling from survey $i$ with a known detection rate, $\mathcal{R}_{i}$, we expect

\begin{equation}
    \mathcal{R}_{\rm DSA} = \mathcal{R}_{i}\,\frac{\Omega_{\rm DSA}}{\Omega_i}\,\left ( \frac{{\rm SEFD}_i}{{\rm SEFD}_{\rm DSA}} \sqrt{\frac{B_{\rm DSA}}{B_i}}\right )^\alpha.
    \label{eq:rate}
\end{equation}

Here, ${\rm SEFD}$ is the system-equivalent flux density, which is the ratio 
of system temperature to gain of the full array. Surveys considered here are assumed to be dual polarization and have similar correlator efficiencies. $B$ is the total usable bandwidth in the FRB search and $\alpha$ is the cumulative source counts power-law index, $\alpha \equiv -\frac{d\ln N(>S)}{d\ln S}$. For a Euclidean distribution (static, non-evolving spectral and luminosity distribution), $\alpha=3/2$. We assume surveys have the same S/N threshold, but note that sub-threshold searching may be possible across the three DSA sub-bands.

The nominal SEFD, bandwidth, and FoV of the CHR/FRB survey sub-bands are shown in Table~\ref{tab:frb_specs}. When forecasting detection rates, we assign priors to each DSA parameter to capture uncertainty in on-sky performance and on parameters of the 
facilities from which we extrapolate. We adopt Gaussian priors $\alpha \sim \mathcal{N}(3/2,\,0.25)$ on the cumulative source-counts index, $\mathcal{R}_{\rm CHIME} \sim \mathcal{N}(10^{3},\,10^{2})\,{\rm yr^{-1}}$ on the CHIME/FRB detection rate, and 10\% priors on the reference-survey sensitivities, ${\rm SEFD}_{\rm CHIME} \sim \mathcal{N}(100,\,10)$\,Jy and ${\rm SEFD}_{\rm HTRU} \sim \mathcal{N}(40,\,4)$\,Jy. The DSA sensitivity carries a larger uncertainty on ${\rm SEFD}_{\rm DSA}$, with a modal value of 1.8\,Jy and a lognormal tail that has $P(>2.0 \rm\,Jy) = 30\%$, reflecting as-yet-unverified on-sky performance and known elevation dependence \citep{flygareDSATsys}. We set a minimum value of 1.65\,Jy for the DSA SEFD. For sub-band 3 we additionally draw the rate spectral index 
from $\gamma \sim \mathcal{N}(0,\,0.25)$ because of the lack of reference surveys at 1.8\,GHz.

\vspace{3mm}
\noindent \textbf{Sub-band 1 (880--980\,MHz):} The lowest CHR/FRB sub-band has the largest FoV ($\sim$\,10.4\,deg$^2$ at 930\,MHz) but the narrowest bandwidth of the three. 
We scale from CHIME/FRB for sub-band 1 because it observes at 400-800\,MHz and its detection rate is well characterized thanks to $>5$ years of finding multiple FRBs per day \citep{frbcollaboration2026secondchimefrbcatalogfast}. 
CHIME/FRB detects roughly $10^3$\,yr$^{-1}$ with an ${\rm SEFD}$ of $\sim100$\,Jy. We take $\Omega_{\rm CHIME} = 200$ \,deg$^2$ and $B_{\rm CHIME}=300$\,MHz, assuming roughly one quarter of the band is lost to RFI. $B_{\rm DSA,1}$ is taken to 
be 100\,MHz, given its placement in a clean region of the spectrum. Computing the rate in Equation~\ref{eq:rate}, we find a CHR/FRB detection rate of 9$\times10^3$\,yr$^{-1}$ in sub-band 1, using the nominal values. The full Monte Carlo results accounting for 
performance uncertainty are shown in Figure~\ref{fig:ratesMC}.

If we instead extrapolate from the DSA-110 (1350-1530\,MHz) detection rate during its commissioning period \citep[e.g.,][]{sharma_preferential_2024}, we obtain a higher predicted value for a flat rate spectral index.
The DSA-110 detected roughly 30 FRBs per year with its first 48$\times$4.65\,m antennas. The DSA's 1650$\times$6.15\,m antennas suggest sub-band 1 will detect roughly $2\times10^4$ FRBs per year, assuming $\alpha=3/2$. This factor-of-two discrepancy could easily be explained by the unknown FRB spectral index between 400--1500\,MHz, different search completeness at different observatories, the RFI environment, or myriad other systematics that arise in practical FRB detection systems.

\vspace{3mm}
\noindent \textbf{Sub-band 2 (1310--1510\,MHz):} To estimate FRB detections in the middle search band, we scale from single-pulse surveys that used the Murriyang / Parkes 21\,cm Multibeam receiver \citep[e.g.,][]{bhandari_parkes}. We choose this approach because of the overlapping frequency coverage, the offline search that enabled Murriyang to characterize its rate, and the high sensitivity of Murriyang. 
Large differences in the sensitivity ratio, $\frac{{\rm SEFD}_i\,\sqrt{B_{\rm DSA}}}{{\rm SEFD}_{\rm DSA}\sqrt{B_{i}}}$, increase dependence on the assumed source counts power-law index $\alpha$. For example, small deviations from the Euclidean value of $3/2$ would lead to large errors if we chose, say, ASKAP in Fly's Eye mode \citep{Bannister_2017}, where point source sensitivity is 1000$\times$ lower than that of DSA. 

Murriyang detected 9 FRBs in 2512\,hrs of observing in its high-lat High Time Resolution Universe (HTRU) single-pulse survey \citep{champion2016}. We take $\mathcal{R}_{\rm HTRU} = 31^{+15}_{-10}$\,yr$^{-1}$. The collecting area of Murriyang is $\sim$\,15 times smaller than that of the DSA. We take ${\rm SEFD}_{\rm HTRU}\approx 40$\,Jy. The usable bandwidth of 
HTRU was roughly 340\,MHz centered at 1352\,MHz. Its FoV is reported as 0.55\,deg$^2$. Extrapolating from Murriyang with Euclidean source counts, 
we find that sub-band 2 on the DSA will discover 1.5$\times10^4$ FRBs per year. The full rate posterior is shown in Figure~\ref{fig:ratesMC}, including uncertainty on $\mathcal{R}_{\rm HTRU}$ and CHR/FRB performance. 

\vspace{3mm}
\noindent \textbf{Sub-band 3 (1650--1950\,MHz):} The high-frequency 
sub-band requires the least compute and memory to search because 
dispersion delays are shorter, scattering is reduced, and 
intra-channel dispersion smearing is negligible at higher frequencies \citep{connor2019}. This means more resources can be devoted to 
high DMs, short timescales, and/or number of beams. However, 
blind FRB surveys have not been carried out at high frequencies, 
so the intrinsic FRB rate above 1.4\,GHz is not well characterized. 
As with sub-band 2, we choose to extrapolate from Murriyang HTRU. In this case, we include a Gaussian prior on rate spectral index, $\gamma$ where $\frac{dn}{d\nu} \propto \nu^\gamma$. The prior is centered on 
zero with a standard deviation of 0.5. This effectively alters the DSA sub-band 3 sensitivity as, 

\begin{equation}
    \frac{\sqrt{B}}{{\rm SEFD}_{\rm DSA}} \rightarrow \left ( \frac{\sqrt{B}}{{\rm SEFD}_{\rm DSA}}\right ) ^{1 + \gamma}.
\end{equation}

The maximum likelihood rate 
in sub-band 3 is $9\times10^3$\,yr$^{-1}$ for $\gamma=0$. The posterior is wider than other sub-bands due to spectral index uncertainty. 

\subsection{Total FRB detection rates}
If all FRBs were narrowband with $\Delta\nu$ below the spacing 
of the three CHR/FRB sub-bands, then one could sum detection 
rates across S1, S2, and S3 to estimate the total detection rate.
While repeating sources are known to show banded structure, most FRBs are relatively narrow in time and broad in frequency \citep{Pleunis_2021b, frbcollaboration2026secondchimefrbcatalogfast}. Summing detections across sub-bands would therefore double count many FRBs. But even for broadband bursts, deviation from flat spectral index, frequency-dependent sensitivity and RFI, and the disparate computational load of each sub-band imply that many FRBs will not appear in all three sub-bands simultaneously. The strongest effect is non-overlapping FoVs: At 930\,MHz the primary beam is nearly $4\times$ larger in area than the beam in the highest sub-band. Even with identical sensitivities and flat FRB flux densities between 0.88-1.95\,GHz, most FRBs detected in sub-band 1 will not be in the FoV of sub-band 3. We note that three independent sub-band searches will enable sub-threshold detection limits. For example, a $6.5\,\sigma$ event at the same sky position, DM, and arrival time detected in multiple bands 
could be trusted, boosting the effective detection rate by 25-75$\%$.

For narrowband FRBs, the
total detection rate is simply the sum of detections in all three bands. This is an upper limit. If all sources are broadband and spectral sensitivity is flat in both single-pulse search and $T_{sys}$, then the total rate is that of sub-band 1 because its primary beam is a superset of the higher bands. Formally, the total rate is bounded by 
the maximum across the three bands and their sum,

\begin{equation}
    \max_i \, \mathcal{R}_i \leq \mathcal{R}_{tot} \leq \mathcal{R}_{S1} + \mathcal{R}_{S2} + \mathcal{R}_{S3}
\end{equation}

We plot the total annual FRB detections in Figure~\ref{fig:ratesMC} in gray. This comes from marginalizing over a flat prior across the bound on $\mathcal{R}_{tot}$. 

The existence of repeaters implies that the total detections will always be higher than the total number of sightlines. Maximizing the number of sightlines enables 
multiple science cases, from FRB cosmology to population studies. For example, repeat pulses from the same FRB carry no new information on the column of cosmic baryons inferred from DM. We wish to estimate the fraction of detected pulses that belong to unique sightlines. 

In CHIME/FRB Catalog 2, roughly 20$\%$ of detected extragalactic pulses were from known repeaters even though fewer than 5$\%$ of sources were repeaters \citep{frbcollaboration2026secondchimefrbcatalogfast, chime_rn4}.
Most CHIME sources spent 100--300 hours in the primary beam 
in five years of observing. The DSA's primary beam is 
$\sim$\,40$\times$ smaller than that of CHIME, so sources will receive less total exposure over the course of its 5-year all-sky survey. With a $\sim$\,5\,deg$^2$ FoV and a $\sim$\,30,000\,deg$^2$ survey footprint, a given FRB will be observed just 0.016$\%$ of the time, decreasing the fraction of FRBs observed as repeaters. There are exceptions, such as possible deep drilling fields and pulsar timing fields, where we expect a larger fraction of repeaters to emerge. We estimate that 85--95$\%$ of pulses detected by 
the DSA will be along unique sightlines based on the lower fractional exposure time compared with CHIME. We expect that less than a few percent of unique sources will be repeaters. Observed repeating CHIME/FRB sources have a shallower DM-implied-redshift distribution than non-repeaters because 
at lower-$z$ a larger fraction of the repetition luminosity function 
will sit above the flux threshold. The DSA will probe deeper into a repeater's luminosity function at a fixed distance, compared to CHIME/FRB. We expect that DSA repeaters will be preferentially closer than DSA non-repeaters.

In forecasting detection rates, we have assumed computing infrastructure that can accommodate forming and searching six million beams at 1\,ms sampling. Details on the search system and early benchmarking will be presented in a Chronoscope instrument paper. If the real-time pipeline cannot perform the FRB search with parameters presented in Table~\ref{tab:frb_specs}, the full FoV could still be searched by tapering in the $uv$ plane to form fewer, larger synthesized beams, reducing sampling time or frequency resolution, or changing the DM plan. We estimate that a search with 10$\times$ fewer beams (i.e., 10$\%$ of the expected compute) would still produce $\sim10^{4}$ FRBs per year combined over the three sub-bands, localized to $\lesssim0.250''$. This is derived from the approximately $2.1\times$ sensitivity reduction of a Gaussian taper to $\sim$\,9\arcsec~with the DSA baseline distribution.

\vspace{3mm}

\subsection{FRB redshift distribution}
The existing sample of localized FRBs spans a wide range of distances, from a few Mpc \citep{bhardwajM81, KirstenM81} to a redshift $z>2$ \citep{caleb2025fastradioburst3}, with a median $z$ of $0.25-0.30$. The observed redshift distribution is non-trivial to model because it is impacted by numerous selection effects, including in host-galaxy follow up and bias against detecting both high- and low-DM events in FRB search. While we cannot predict distances with precision, is likely that 
DSA's sensitivity will result in a deep redshift 
distribution and a relative abundance of high-DM events compared to current surveys. Naively, the mean luminosity distance, $\left < d_L\right >$, of a survey sample scales as the inverse square root of the flux threshold, $\frac{1}{\sqrt{S_{min}}}$, obeying the inverse square law \citep{dongzi_2018}. However, this requires a 
non-evolving population and high completeness in the FRB pipeline. 
For example, CHR/FRB has a $\sim60\times$ lower detection threshold than DSA-110, putting the mean CHR/FRB event at $7\times$ higher luminosity distance than DSA-110. But incompleteness at high DMs 
or a decline in the volumetric rate at high-$z$ would break the inverse square law assumption of $\left < d_L \right > \propto S_{min}^{-1/2} $.

We estimate the DSA FRB redshift distribution under two simplified
models for the intrinsic population. In the first, the comoving volumetric
rate tracks the cosmic star-formation history, for which we adopt the
\citet{madau} form,

\begin{equation}
    \psi(z) \propto (1+z)^{2.7} / \left[1 + \left((1+z)/2.9\right)^{5.6}\right].
\end{equation}

In the second, FRBs trace stellar mass, which skews the detected sample toward
lower redshifts. In both cases the observed distribution is the intrinsic
rate multiplied by the comoving volume element $\mathrm{d}V/\mathrm{d}z$ and
by a survey selection function \citep{Macquart_2018}. We assume the luminosity function is a power
law that does not evolve with redshift, with differential slope
$\gamma \approx -1.9$ as inferred from ASKAP/CRAFT and Murriyang samples
\citep[e.g.,][]{james2022}, and encapsulate the resulting fluence-limited
completeness in a smooth sensitivity rolloff,
$s(z) = \left[1 + (z/z_0)^{4}\right]^{-1}$ with $z_0 = 1.3$ chosen to match
the expected DSA horizon for the fiducial energy function. The effective observed 
redshift dependence in the stellar mass case is roughly, $\rho_\star(z) \propto e^{-z/0.45}$.

\begin{figure}[htbp]
    \centering
    \includegraphics[width=\textwidth]{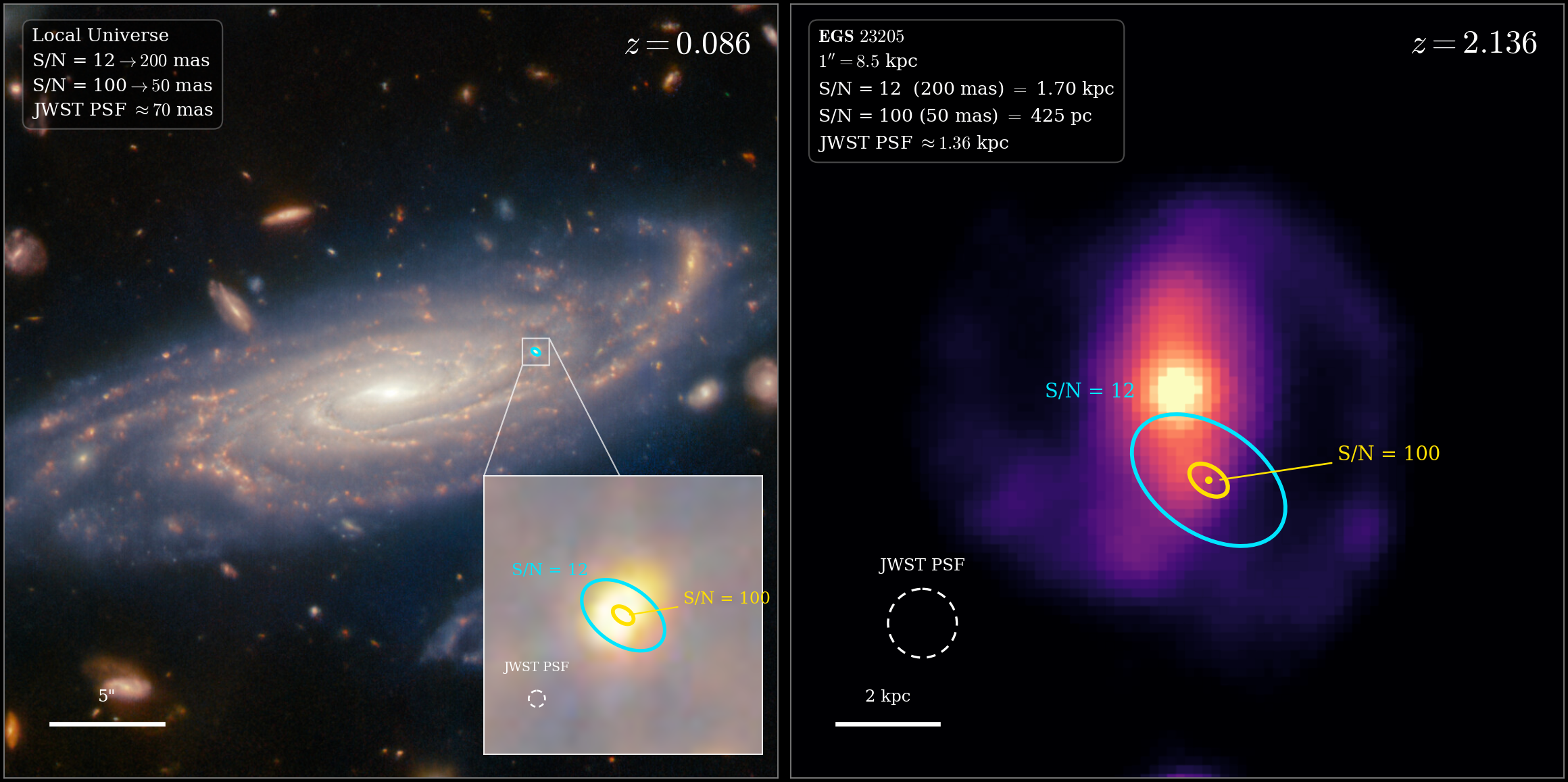}
    \caption{An illustrative example of two host galaxy localizations that are possible with the CHR/FRB survey. The left panel shows a low-redshift galaxy (LEDA\,2046648, $z\approx0.10$) for a typical FRB localization (S/N$=12$, $\theta_{\rm FRB}\approx0.2''$, cyan ellipse) and a bright burst (S/N=$100$, $\theta_{\rm FRB}\approx0.05''$, yellow ellipse). 
    The LEDA\,2046648 image comes from public JWST/NIRCam commissioning data. The right panel shows a $z\approx2$ galaxy and the corresponding normal and bright burst error 
    regions. This image was obtained by the JWST/CEERS survey using NIRCam F444W. For nearby or high-SNR events, the physical localization will enable precise offsets and studies of the immediate FRB environment.
    \vspace{5mm}}
    \label{fig:hostgalaxy}
\end{figure}

\subsection{Localization precision}
\label{subsec:localization_precision}

At zenith, the DSA PSF will have a mean FWHM of $\sim$\,2.4'' at 1400\,MHz. 
Astrometric precision ($1\sigma$) for transient point sources such as FRBs is $\theta_{\rm FRB} \approx 0.45\times \frac{2.4''}{S/N}\,\left ( \frac{1400\,\rm MHz}{\nu} \right )$ \citep{reid_astrometry}, 
with a floor at high-S/N determined by PSF modeling, reference frame precision, and other systematics. The $90\%$ confidence region is roughly $2.14\,\theta_{\rm FRB}$. With a 
detection threshold of 8\,$\sigma$, we expect the median CHR/FRB event 
to be localized at 90$\%$ confidence to $\sim300$\,mas in S1, $\sim190$\,mas in S2, and 
$\sim150$\,mas in S3, assuming Euclidean source counts. Based on the rate estimates in Section~\ref{sec:forecasts}, we find that DSA should find 
an FRB with S/N\,$\geq60$ every day, resulting in a localization precision of roughly $\leq$\,50\,mas. For most FRBs, we expect host galaxy identification to be unambiguous \citep{Eftekhari_hosts}. Hundreds of bright bursts per year could be localized to $\leq225\,\mathrm{pc}\,\left (\frac{z_{host}}{0.30} \right )$.

The CHIME/FRB baseband localization pipeline achieves significant super-resolution \citep{chimebase}, with a systematics-limited localization floor at S/N$\,\approx30$ in large part due to their strongly direction-dependent and difficult-to-model beam phase.
The CHIME/VLBI Outriggers do not currently super-resolve the majority of CHIME-detected FRBs due to residual ionospheric calibration errors.
The reduced collecting area of the CHIME/FRB Outriggers relative to the core CHIME array also significantly limits the interferometric fringe S/N for the bulk of the detected FRB population \citep{chimevlbiloc, chimevlbi}.
The 15 outriggers of DSA-110 can achieve a factor of 15-20 in super-resolution.
There is reason to believe the DSA's astrometric super-resolution 
will be less affected by systematics than existing FRB surveys.
The large number of DSA antennas organized in a pseudo-random distribution leads to dense $uv$-coverage and an excellent PSF that will be well characterized \citep{hallinan2019, polish}. The DSA observing frequencies are higher compared to CHIME/FRB, which means ionospheric delays will be a smaller systematic.
The Radio Camera backend will continuously image the same sky
as the FRB survey, producing an astrometric WCS solution for each 21\,minute pointing. The mean DSA pointing will have $\mathcal{O}(5)$ 
VLBI calibrators above 1\,mJy as well as hundreds of NVSS sources detected at $\geq1000\,\sigma$. The DSA will conservatively have a 50\,mas precision in the reference frame solution; we take that as our lower-limit on FRB localization. In Figure~\ref{fig:hostgalaxy} 
we visualize the DSA's FRB localization ellipses on top of two JWST/NIRCam galaxies. We show a median S/N=12 FRB (cyan ellipse) as well as a S/N=100 FRB.

\subsection{Host galaxy follow up}
Most FRB science cases require identifying a host galaxy and obtaining its redshift. With sufficiently deep imaging, hosts should be clear-cut with DSA's localization precision \citep{Eftekhari_hosts, PATH, james2026b}. Follow up of current samples ($\mathcal{O}(100)$ FRBs) has been done by 
small groups observing individual FRBs with optical and infrared spectroscopic instruments. The size 
of the sample produced by the CHR/FRB survey demands a broader approach to host galaxy characterization. This approach must incorporate public spectroscopic galaxy surveys, photometric redshifts, 21\,cm derived redshifts, and mm/sub-mm instruments in addition to targeted host follow up (see Figure~\ref{fig:allsky}). In Table~\ref{tab:optsurveys} we list nine surveys that could provide host galaxy follow up. For some science cases, including FRB cosmology, it may be possible to model the redshift distribution as measured from a subset of hosts \citep{reischke26a}, alleviating the need to obtain redshifts of all sources.

\begin{deluxetable}{lcccc}
\tabletypesize{\footnotesize}
\tablecaption{Surveys for DSA FRB host identification and redshifts. \label{tab:optsurveys}}
\tablehead{
\colhead{Survey} & \colhead{Bands} & \colhead{$f_{\rm sky}$} &
\colhead{Depth (AB)} & \colhead{Data}
}
\startdata
Rubin/LSST & $ugrizy$          & 0.3 & $r\simeq27.5$ & photometry \\
Argus Array     & $g$, $r$          & 0.9 & $g\simeq27$   & photometry \\
Euclid     & $I_{\rm E}$, $YJH$ & 0.3 & $\simeq$26.2  & photometry \\
Roman      & $YJH$             & 0.1 & $\simeq$26.5  & photometry \& grism \\
SPHEREx    & 0.75--5.0\,$\mu$m         & 1.0 & $\simeq$19.5  & spectro-photometry \\
NANCY\tablenotemark{a} & F146  & 1.0 & $\simeq$25.5  & photometry \\
DESI       & 0.36--0.98\,$\mu$m        & 0.5 & $r<20.2$      & spectroscopic \\
Via/Boombox        & 0.36--1.0\,$\mu$m        & \nodata & $r<24$  & spectroscopic \\
SDSS-V     & 0.36--1.0\,$\mu$m         & \nodata & $r\lesssim22$ & spectroscopic \\
DSA/21\,cm        & $0.7-1.4$\,GHz        & 1 & \nodata & Radio IFU 
\enddata
\tablecomments{$f_{\rm sky}$ is the fraction of the DSA sky (Dec $>-30\arcdeg$)
covered; Via and SDSS-V are targeted fiber programs. Depths are $5\sigma$
final-coadd point-source magnitudes (BGS Faint limit for DESI).}
\tablenotetext{a}{Proposed all-sky survey with Roman.}
\vspace{3mm}

\end{deluxetable}

\vspace{3mm}

\noindent\textit{Imaging:} 
The combination of Rubin/LSST, Argus Array, Euclid, and Roman will be transformative for FRB host galaxy identification. Rubin/LSST 
is expected to detect $\sim$\,$10^{10}$ galaxies down to 
$r < 27.5$ by 2036 \citep{ivezic}. The redshift range will be $0<z<3$ with a median redshift of roughly 1. A recent study by \citet{James2026a} found that 
10 year co-added images would identify 81\% of FRBs discovered in MeerTRAP's coherent mode, which is the nearest in sensitivity to DSA. The DSA footprint extends down to Dec\,$\gtrsim-30^\circ$, so sky coverage overlap will depend on how far north LSST ultimately surveys. The current expectation is that LSST's Wide-Fast-Deep survey will reach Dec$\sim+15$\footnote{https://survey-strategy.lsst.io/baseline/wfd.html}, providing more than 10,000\,deg$^2$ of overlap with DSA. 

The Argus Array will deliver deep optical imaging over most of the DSA footprint, including the large northern area outside Rubin's Wide-Fast-Deep survey \citep{Law_2022}. The planned array of approximately 1,200 0.28-m telescopes will operate as an 8-m-class instrument, imaging 8,000\,deg$^2$ simultaneously with approximately 2'' delivered image quality. Its baseline two-band survey will cover Dec\,$\gtrsim-20^\circ$, overlapping more than 27,000\,deg$^2$ of the DSA survey, and is expected to reach $g\sim24.7$ in weekly coadds and $g\sim26.5$ after six months, with multi-year coadds approaching $g\sim27$. Argus will therefore provide deep imaging for DSA-localized FRBs, identify host candidates substantially fainter than those in existing all-sky surveys, and supply colors and magnitudes for host-association analyses and spectroscopic target selection \citep{freeburn26}.

Euclid was launched in July of 2023 and will obtain Visible imaging at 0.17'' resolution and a depth of $m_{\rm AB, 5\sigma}\sim26.2$ and NIR imaging down to $\sim$\,24.5 \citep{euclid}. Nearly 10,000\,deg$^2$ of Euclid's total 14,000 \,deg$^2$ footprint will overlap with the CHR/FRB survey. 
Roman will provide deep, Hubble-quality near-infrared imaging of DSA-localized FRB hosts in the common footprint, enabling secure host identification. As with Euclid, Roman's image fidelity will enable detailed measurements of host morphology, size, stellar mass, dust attenuation, and star-formation history. DSA/CHR's localization precision will identify FRBs in spiral arms and knots of star formation. Roman's High-Latitude Wide-Area Survey will obtain $YJH$ imaging and slitless grism spectroscopy over approximately 2,400\,$\mathrm{deg}^{2}$ to a characteristic $5\sigma$ point-source depth of $m_{\rm AB}\simeq26.5$. An additional $2,700\,\mathrm{deg}^{2}$ will be imaged in the $H$ band to $m_{\rm AB}\simeq26.2$, yielding more than 5,000\,$\mathrm{deg}^{2}$ of Hubble-resolution near-infrared imaging \citep{roman2025}. The  Next-generation All-sky Near-infrared Community surveY (NANCY) on Roman has been proposed to image the entire sky at $\sim$\,0.1'' resolution 
in a single band down to $H\sim25.5$ \citep{Han2026}. This would benefit the DSA FRB survey and the broader FRB community greatly. We show a simulated sky distribution of DSA FRBs in Figure~\ref{fig:allsky} along with the footprints of upcoming imaging surveys. 

Existing all-sky imaging surveys such as Pan-STARRS, the DESI Legacy Imaging Surveys, WISE, and others, will continue to 
play an important role in host identification. 

\begin{figure}[htbp]
    \centering
    \includegraphics[width=0.825\textwidth]{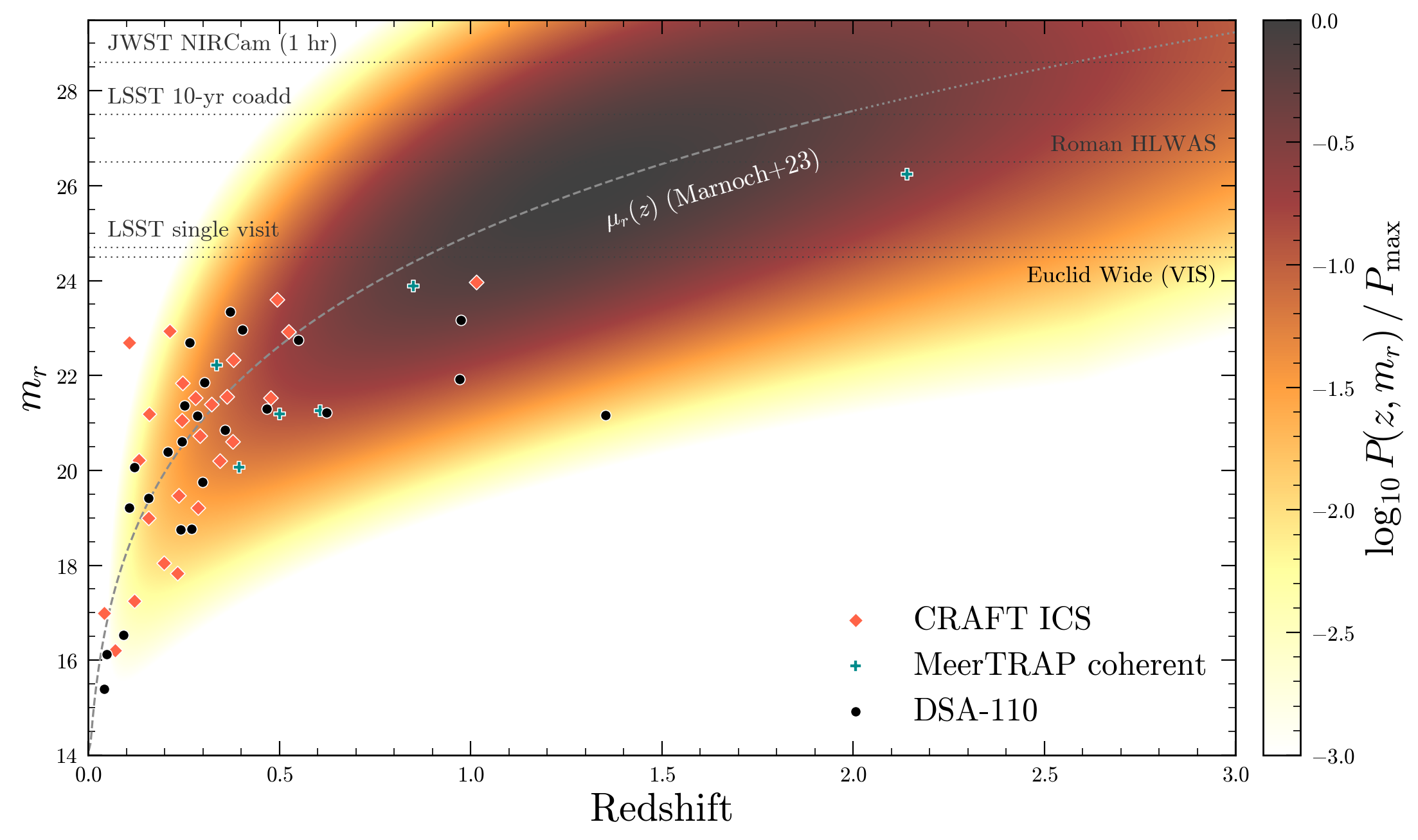}
    \caption{The $r$-band magnitude/redshift distribution of three current FRB samples (ASKAP/CRAFT, MeerKAT/MeerTRAP, and DSA-110) and a simulated DSA CHR/FRB host distribution.
    There already exists incompleteness for faint, high-$z$ hosts in the observed FRB sample. For the CHR/FRB heatmap, we assume FRB rates trace cosmic star formation, skewing the DSA distribution to higher redshifts due to its sensitivity. Following \citet{James2026a}, we show the LSST single visit and 10-year coadd limits on $m_r$.}
    \label{fig:host_mag_redshift}
\end{figure}

In Figure~\ref{fig:host_mag_redshift}, we estimate the $m_r$/redshift relation for 
FRB host galaxies in the DSA sample. 
We plot existing FRB redshifts from CRAFT, MeerTRAP, and DSA-110, with limits from 
aforementioned upcoming optical surveys. The simulated DSA host distribution is the joint density
$P(z, m_r) = p(z)\,\mathcal{N}\!\left(m_r;\, \mu_r(z), \sigma_r(z)\right)$.
The redshift distribution $p(z)$ assumes the FRB rate traces the cosmic
star-formation history, modulated by the comoving
volume element and a smooth sensitivity rolloff at $z \sim 1.3$ assumed for
the DSA in Section~\ref{sec:forecasts}. The true selection function will ultimately be set by the survey's
joint fluence--DM--width completeness. For the host magnitudes we adopt the
empirical FRB host magnitude--redshift relation of \citet{James2026a},
which is anchored to the spectroscopic sample of
\citet{marnoch23}. We extrapolate $\mu_r$ smoothly beyond $z = 2$, neglecting galaxy evolution as
in the original work \citep{James2026a}.

\vspace{3mm}

\noindent\textit{Photometric and spectrophotometric redshifts:} SPHEREx will provide homogeneous, all-sky near-infrared (NIR) spectrophotometry, 
delivering as many as 450\,M galaxies with $\sigma_z < 0.1\,(1+z)$ 
and 20\,M with  $\sigma_z < 0.003\,(1+z)$ \citep{spherex_redshift}.
All DSA FRB host galaxies will be in the $4\pi$\,sr SPHEREx footprint, 
but many will be too faint to be detected ($m_{\rm AB,5\sigma} \gtrsim 19.5$ in the near-IR) or blended with other sources in the 6\arcsec~SPHEREx pixels. Still, the uniform NIR sky coverage will be valuable for characterizing bright hosts outside existing spectroscopic footprints, for prioritizing uncertain or distant hosts for deeper follow-up observations, and for identifying surrounding large-scale structure.

As with host identification in imaging, Rubin/LSST, Euclid, and Roman will enable percent-level photometric redshifts for billions of galaxies \citep{ivezic,euclid}. The DESC Redshift Assessment Infrastructure Layers (RAIL) framework is an open-source suite of modular tools to ingest multi-survey photometry and produce redshift posteriors \citep{RAIL}. The CHR/FRB will employ RAIL and 
similar tools to account for asymmetric errors and marginalize over catastrophic photo-$z$ failures that can impact FRB cosmology and population studies. 

\vspace{3mm}

\noindent\textit{Spectroscopic redshifts:} The Dark Energy Spectroscopic Instrument (DESI) \citep{desi2016} has now obtained 
spectra for nearly 50\,M galaxies and quasars\footnote{https://www.desi.lbl.gov/}. 
Approximately half of DSA FRBs (Dec $\gtrsim$ -30$^\circ$) will overlap with the DESI footprint. 
The four principal DESI tracer samples will serve as potential host galaxies and excellent foreground catalogs against which to correlate FRBs. These include the Bright Galaxy Survey \citep[BGS, $z\lesssim0.4$;][]{BGS}, 
Luminous Red Galaxies \citep[LRG, $0.3 \lesssim z \lesssim 1.0$;][]{Zhou_2023}, 
Emission-Line Galaxies \citep[ELG, $0.6 \lesssim z \lesssim 1.6$;][]{ELG}, 
and even the quasar sample ($0.6 \lesssim z \lesssim 1.6$ and $z>2$ for Ly$\alpha$ forest studies). We estimate that 5-15$\%$ of $z<0.5$ DSA host galaxies will belong to the DESI Bright ($r<19.5$) or BGS Faint ($r<20.175$) redshift samples, accounting for the footprint. Depending 
on the FRB luminosity function and $\frac{dn}{dz}$, this could amount to $\mathcal{O}(10^3)$ public host galaxy spectra and redshifts. The ELG sample at higher redshifts is typically blue, star-forming galaxies with strong O\textsc{ii}, with a density of 2,400 ELG candidates per square degree. 
Assuming FRBs trace star formation, these galaxies may be overrepresented as hosts at $z\approx1$. While the DESI LRG sample will 
be invaluable for cross-correlation studies, it remains unclear what fraction of FRBs reside in massive early-type galaxies \citep{Eftekhari_queiscent}
and therefore how many DSA hosts will be DESI LRGs.

The DSA's spectroscopic HI survey is expected to detect several million galaxies at $0<z<1$,
surpassing all previous HI surveys combined \citep{hallinan2019}. A subset of FRBs and their host galaxies will therefore be characterized in the radio alone, particularly at low redshifts. The all-sky survey will have a limiting detection mass of $M_{\mathrm{H}\textsc{i}} \approx 9\times10^{10}$ at $z=0.20$. In other words, for a Milky Way-like FRB host galaxy, DSA/21\,cm will provide a redshift and kinematics out to $\sim$\,Gpc distances for free. Depending on the final FRB redshift distribution, $1-10\%$ of hosts could be at $z\leq0.30$. We anticipate that hundreds to a few thousand host galaxies could be obtained from DSA $\mathrm{H}\textsc{i}$. Parallel spectroscopic 21\,cm observations should serve to identify star-forming dwarf galaxies with large HI fractions, which are not typically included in cosmological galaxy surveys like DESI. Dwarf galaxies are known to host repeating FRBs that reside in unusually dense magnetoionic environments \citep{chatterjee_direct_2017,Niu2022,moroianu_milliarcsecond_2025}. The FRB/21\,cm synergy has already been demonstrated for host galaxies in the CRAFT/WALLABY (FRB/HI) surveys on ASKAP \citep{Glowacki_2023}.


The Via Project is an all-sky spectroscopic survey on the 6.5m MMT and Magellan/Clay telescopes, currently under construction. Via will total more than 500 nights. Its primary scientific instrument, Viaspec, is an $R\approx 15,000$ stabilized spectrograph with over 500 fibers \citep{theviacollaboration2026projectoverviewscienceinstrument}. Via will also have a low-res spectrograph, Boombox ($R \approx 1,000$) spanning $360-1010$\,nm with 36 fibers. Via/Boombox could deliver $\sim$\,1,000 DSA FRB host galaxies with $z \lesssim 1.3$ and $r <24$, over five years \citep{theviacollaboration2026projectoverviewscienceinstrument}. This would include both spare Boombox fibers on Viaspec fields that have a DSA FRB host galaxy, and targeted pointings.

The open fiber program on SDSS-V provides an opportunity for CHR/FRB host galaxy follow up. For a 1\,hr BOSS exposure on SDSS-V, we estimate that 
redshifts could be obtained for FRB host galaxies with strong emission lines at $m_r \lesssim 22$ \citep{Ross_2012}. Analogous 
programs exist on other multiplexed spectroscopic surveys, such as 
the DESI Secondary Target Programs \citep{DESI_secondary} and Subaru/PFS Community Fillers \citep{PFS}. Each will be assessed based on expected science return and the DSA FRB team's access to these programs. 

The extraordinary imaging and spectroscopic capabilities of JWST make it a compelling FRB follow up telescope. It has proven valuable in this capacity already, with JWST/NIRCam F150W2 images resolving $\sim$\,10\,pc scales in the vicinity of the VLBI-localized FRB\,20250316A \citep{blanchard2025}. The Near Infrared Spectrograph (NIRSpec) was used to obtain a redshift of 
$z=2.148$ for FRB\,20240304B \citep{caleb2025fastradioburst3}. The high-redshift tail of 
DSA FRBs makes JWST even more critical. Limited resources and significant competition for telescope time mean that a tiny fraction of DSA FRBs could be followed up, but JWST will be an essential tool for 
$z\gtrsim3$ bursts and bright, local-Universe events where the Chronoscope will have $\lesssim100$\,pc localization uncertainty (e.g. Figure~\ref{fig:hostgalaxy}). In a similar vein, ALMA is a powerful telescope 
for high-redshift galaxy spectroscopy. 
It can detect bright far-infrared and molecular lines such as [C II] 158\,$\mu$m, [O III] 88\,$\mu$m, CO and [C I].
The ALMA-ALPINE survey detected [C II] in a substantial fraction of ordinary star-forming galaxies at $4.4<z<5.9$ \citep{ALMAalpine}, demonstrating that this approach is not restricted to extreme quasars or submillimeter galaxies. Additionally, the favorable submillimeter K-correction makes dust emission detectable across roughly z\,$\sim1–10$, although intrinsically low-mass hosts will remain difficult.
Like JWST, ALMA time is highly competitive so a carefully selected sample of potential high-$z$ FRBs would be targeted in mm/sub-mm. 

Other dedicated ground-based follow-up of DSA FRBs may be conducted with large facilities, as has been the case for the existing sample of localized events \citep[e.g.,][]{shannon2025craft,sharma_preferential_2024,Jankowski_2023,leung25}. The possibility of using the next generation of highly sensitive multi-object spectrographs and IFUs operating over $\gtrsim$arcminute fields (e.g., GTC/Megara, VLT/MOONS, Keck/LRIS2, ELT/MOSAIC) is particularly important for simultaneous redshift measurements, host-environment characterization, and the characterization of intervening galaxies or structures.

\vspace{3mm}
\noindent 
{\it Propagation based redshifts:}  Redshifts and host-galaxy contributions to DM can be estimated by exploiting the relation between DM and pulse broadening times $\tau$ from scattering.   The $\tau({\rm DM})$ relation can be calibrated using Galactic pulsars and FRBs that have well constrained DM and scattering inventories \citep[][]{2022ApJ...931...88C}.  Large FRB samples with both redshifts and scattering times will yield the hyperparameters for a refined relation that can then be inverted into redshift estimates for the large number of FRBs that will not have measured redshifts.  It is also possible that scattering will be detected from circumgalactic media, particularly at redshifts $z > 1$, which will lead to further refinement of the redshift estimator.

\section{FRB science with the DSA}
\label{sec:frb_science}

In this section we offer a non-exhaustive list of science 
cases that can be done with the DSA CHR/FRB survey. 
We group the science cases by ``FRBs as a tool'' and 
``the physical origin of FRBs''. The former category includes applications to cosmic baryons and feedback, as well as 
cosmology and fundamental physics. To reach their full potential, each of these requires hundreds of times more FRB sightlines than currently exist. This is what the DSA will provide. Addressing the physical origin of FRBs includes understanding their host galaxies, their local environments, and the coherent emission mechanism. Such science often requires a large sample from which to draw unusual events that can be studied in great detail. We focus on areas that are not possible without the large well-localized sample of DSA FRBs spanning the full extent of cosmic history.

\vspace{3mm}

\subsection{FRBs as a probe}

\subsubsection{Cosmic baryons \& astrophysical feedback}
Since their discovery in 2007 \citep{lorimer2007}, FRBs have been recognized as promising probes of the ionized cosmic gas that constitutes the majority of baryons in the Universe. 
Most baryonic matter is not confined to stars or the interstellar medium, but resides in the intergalactic medium (IGM) and 
in dark matter halos. It is diffuse, hot, and ionized, 
making the gas difficult to measure directly, hence the decades-old ``missing baryon problem'' \citep{Persic_1992, fukugita}. FRB dispersion provides 
percent-level measurement of baryon columns to cosmological distances, including Milky Way and host galaxy contributions. The distribution of the ionized baryons in and around dark matter halos has significant implications for astrophysical feedback, which is essential for understanding galaxy formation and evolution. The baryon distribution also acts as a critical 
systematic uncertainty in precision cosmology, limiting 
the ability of weak lensing experiments to measure
the cosmological parameters \citep{Chisari_2019, Amon_2022}.

In the past several years, the first samples of FRBs with host galaxy localizations have been used to constrain the cosmic gas distribution and resolve the missing baryon problem \citep{macquart2020, connor2025}. The DM/redshift distribution of $\mathcal{O}(100)$ sources has been used to constrain feedback strength and the fraction of baryons in the IGM \citep{khrykin, connor2025, reischke26, sharma2026signaturessuppressedmatterclustering}. Recently, cross-correlations between FRB 
DMs and other tracers of the large-scale structure (LSS) have 
produced detections \citep{hussaini25, wang25, sharma2026backlightingcosmicwebfast, tszfrb, mccarty2026b}. Finally,
local universe FRBs that have little DM contribution from the 
IGM have been used to directly constrain the CGM baryon content, both of the Milky Way and nearby host galaxies \citep{cook23, ravi25, leung25, mccarty2026a}. 

Each of these methods is limited by the total number of 
FRB sightlines. The CHR/FRB sample will enable 
large cross-correlation studies 
of DM against foreground galaxies, the thermal Sunyaev–Zeldovich effect (tSZ), X-ray maps, void catalogs, Ly$\alpha$ surveys, weak lensing, and cluster catalogs \citep{sharma2026backlightingcosmicwebfast}. We expect high-significance detections for all of these cross-correlations. 

These two-point statistics offer complementary but distinct information about the relationship between cosmic baryons, feedback, and the LSS. For example, DM$\times$tSZ and DM$\times$galaxies will reveal the fraction of non-thermal pressure at the outskirts of dark matter halos, offering insights into the role of cosmic rays and magnetic field pressure \citep{medlock26}.  
The cross-power spectrum of DMs with weak lensing shear maps will be another powerful measurement \citep{reischke26a,leungWL,Wayland_2026}. Weak lensing surveys are sensitive to the total matter integrated along the line of sight; FRB DMs measure only the baryons. Combined, weak lensing surveys can be calibrated and alleviated of their primary systematic: baryonic feedback. By including the DSA sample, FRBs can significantly improve weak lensing constraints on the sum of neutrino masses, $\Omega_b$, and the primordial slope of the power spectrum, $n_s$ \citep{reischke26a}.

By combining DM/galaxy cross-correlation measurements with the kinematic Sunyaev–Zeldovich effect (kSZ), the galaxy-electron power spectrum, $P_{ge}(k)$ can be measured across a wide range of physical scales \citep{Madhavacheril_2019}. This method was recently demonstrated by \citet{mccarty2026b} using localized FRBs and recent kSZ stacks on DESI galaxies \citep{Hadzhiyska25}. 

\begin{figure}[htbp]
    \centering
    \includegraphics[width=0.95\textwidth]{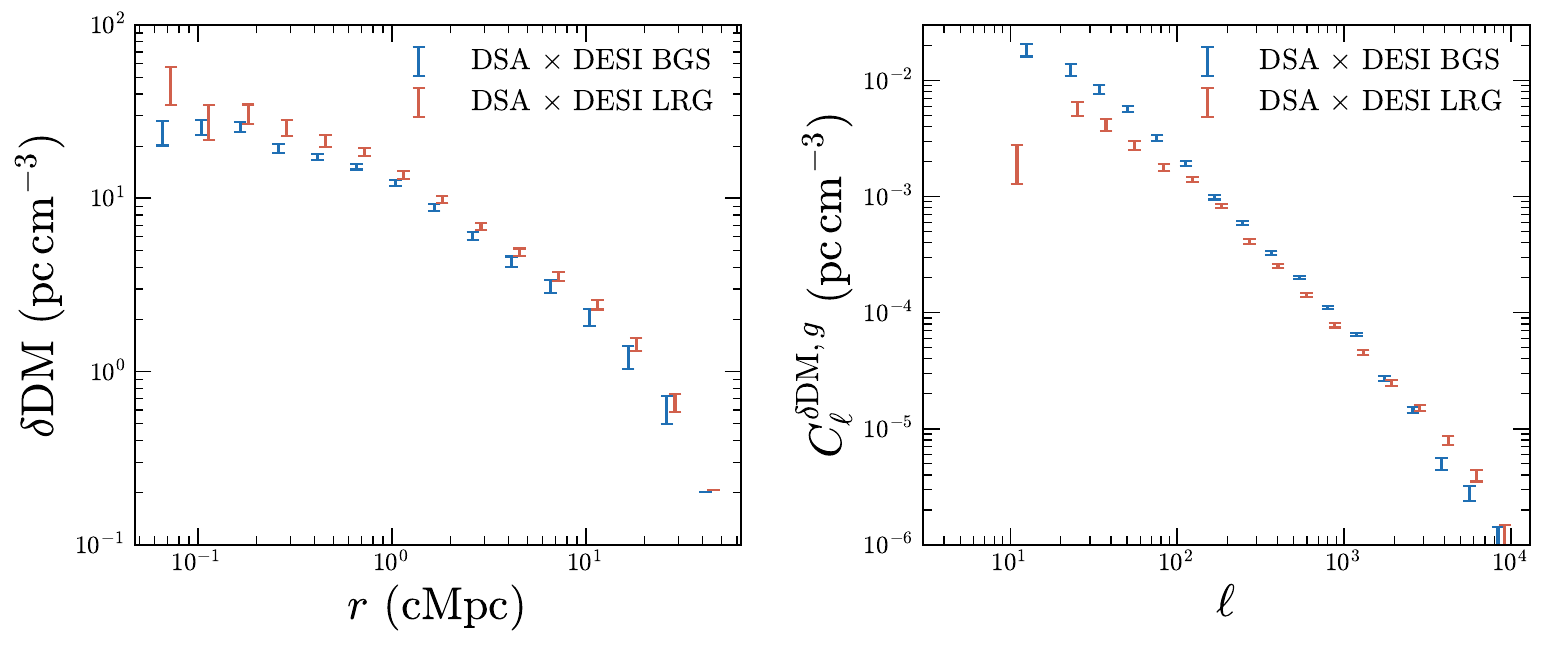}
    \caption{We show forecasted constraints on excess DM vs. transverse scale (left)  and the DM/galaxy angular power spectrum (right) from cross correlating 50,000 DSA-like FRBs with foreground galaxies. We correlate ray-traced DMs from FLAMINGO L1\_m9 with DESI-like galaxies in the simulation. Constraints on the baryon distribution are shown from 60\,kpc (CGM-scale) to 50\,Mpc.}
    \label{fig:baryon_powerspectrum}
\end{figure}

The DM$\times$galaxies statistic measures the circumgalactic media and gas in correlated large scale structure around galaxies, making it a sensitive probe of baryonic physics. With only $\sim100$ sources, this correlation already provides useful insights into feedback \citep{mccarty2026b}. The full DSA sample will provide precision constraints across orders of magnitude in scale and halo mass, which are hard to access with other probes. In Figure \ref{fig:baryon_powerspectrum}, we show forcasted constraints on the DM$\times$galaxies correlation in both configuration and harmonic space for a DSA-like sample. The forecast is made following the procedure outlined in \cite{mccarty2026b}, using DM and halo lightcones from the default L1\_m9 FLAMINGO simulation \citep{flamingo8sigma}. We assume 50,000 DSA FRBs at z=1, and use an abundance matching technique to select samples of simulated subhalos that resemble the DESI photometric BGS \citep{BGS} and LRG \citep{Zhou_2023} samples.

\begin{figure}[htbp]
    \centering
    \includegraphics[width=0.97\textwidth]{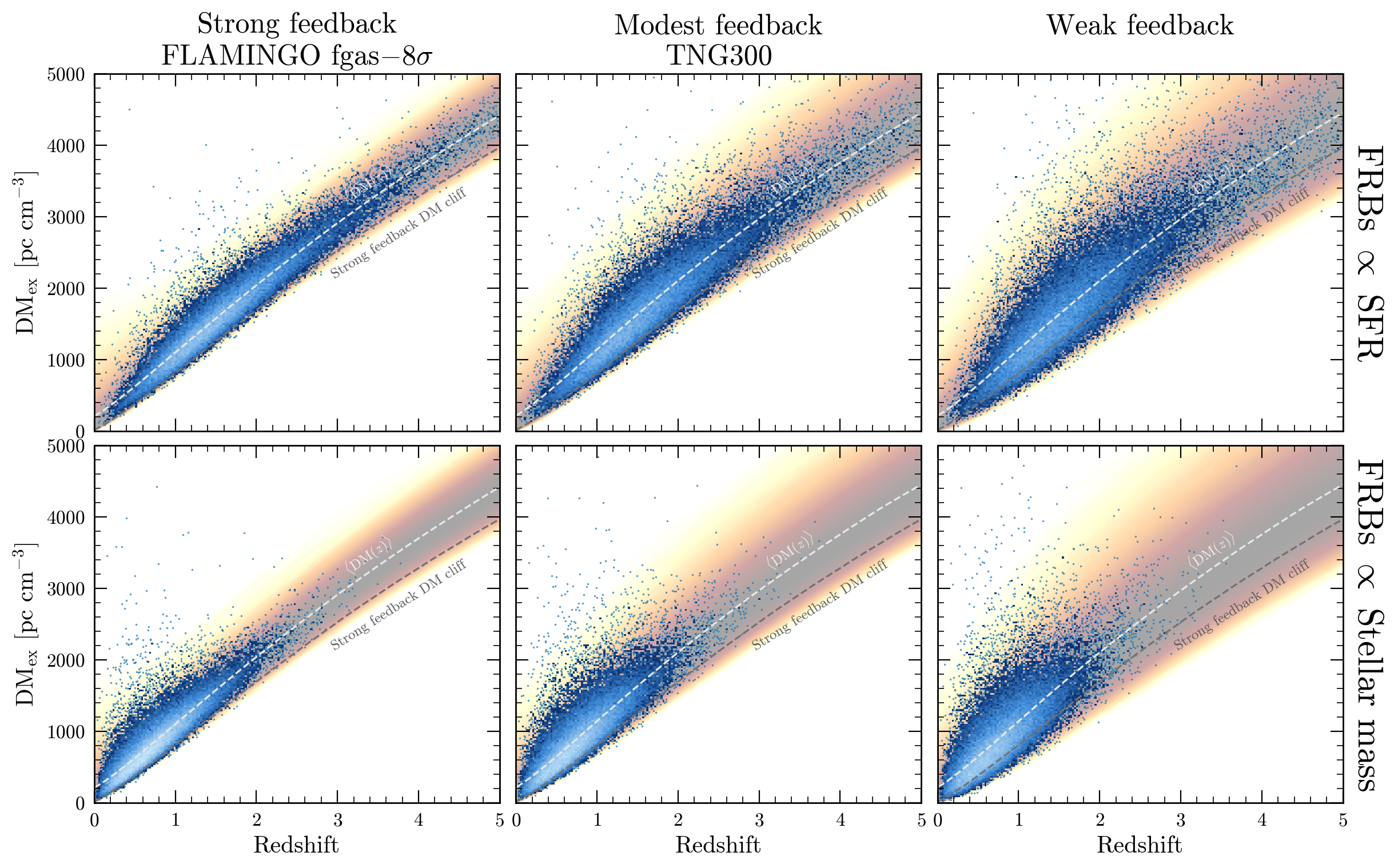}
    \caption{Simulated distributions of extragalactic DM and redshift for different redshift evolution and feedback scenarios. The blue scatter-density markers show 50,000 synthetic FRBs observed with the DSA. The background heatmap is a log PDF of total extragalactic DM at a given redshift, including a lognormal host DM distribution. 
    The left column corresponds to the strong feedback scenario in FLAMINGO fgas$-8\sigma$ \citep{flamingo8sigma}. The middle columns are Illustris/TNG300 DMs from \citet{konietzka} 
    and the right columns are an ad hoc model in which cosmic DM variance is 2$\times$ larger than TNG300. The gray dashed curve corresponds to the ``DM cliff'' in the stronger feedback scenario, taken here to mean the DM value at which $P(<\mathrm{DM_{ex}} | z) = 0.01$.
    \vspace{5mm}}
    \label{fig:dmzgrid}
\end{figure}

In addition to two-point statistics, the DM/redshift distribution of FRBs contains 
information about the distribution of cosmic baryons \citep{macquart2020, connor2025, sharma2026signaturessuppressedmatterclustering}. In Figure~\ref{fig:dmzgrid}, 
we show $P(\mathrm{DM_{ex}}\,|\, z)$ 
for three feedback scenarios: the FLAMINGO $f_{gas}-8\sigma$ simulation (left column, 
strong feedback, \citep{flamingo8sigma}), the Illustris/TNG300 simulation (middle column, modest feedback, \citep{konietzka}),
and an ad hoc 
lower-feedback scenario in which sightline variance of TNG300 DMs is increased by a factor of 2. The FLAMINGO $f_{gas}-8\sigma$ DMs were computed from public lightcone DM maps at $z<3$ and extrapolated to $z=5$. Each feedback model predicts different per-redshift DM variance and ``DM cliffs'' \citep{james2022b}. The DM cliff is the statistical minimum DM (e.g. $P(\mathrm{< DM} | z) = 0.01$) that encodes the baryon fraction in the IGM and the overall distribution of gas at large scales. It is very sensitive to feedback strength \citep{connor2025}.

The discovery of nearby FRBs or extragalactic pulsars by the DSA will provide a gas column that is roughly,

\begin{equation}
\mathrm{DM_{obs}}(z < 0.1) \approx \rm DM_{ISM, MW} + DM_{CGM, MW} + DM_{ISM, Host} + DM_{CGM, Host},
\end{equation}

\noindent because the IGM is subdominant for sources within a few hundred Mpc \citep{mccarty2026a}. Low-luminosity pulses that are undetectable by existing surveys will therefore be incisive for determining the baryonic mass 
in the CGM, if $\rm DM_{ISM, Host}$ can be modeled. For example, a 
dispersed pulse from Andromeda would measure the shared plasma in the Local Group's dark matter halo, and inform claims of a ``baryon bridge'' connecting our Galaxy with M31 \citep{M31bridge}.

\vspace{3mm}

\noindent \textit{HeII and HI reionization}: Thanks to its sensitivity and high $\rm DM_{max}$, 
we expect the DSA to discover a sample of FRBs at $z>2$. 
If FRBs are produced beyond the peak of cosmic star formation, 
DSA should discover sources out to Helium reionization and even 
Hydrogen reionization. FRB DMs for these epochs provide a reionization optical depth and information about IGM evolution \citep{Heimersheim_2022}. The maximum cosmic DM is expected to be $\sim$\,5\,$\times 10^{3}$\,pc\,cm$^{-3}$ at which point the mean DM should plateau in redshift. \citet{beniamini} argue that $\sim$\,40 sources at $6<z<10$ would provide a reoinization optical depth, $\tau_{\rm T}$, measurement that is competitive with Planck. In the absence of host galaxy redshifts, a ceiling on cosmic DM would also measure $\tau_{\rm T}$. Constraining $\rm DM_{max, cos}$ to better than 500\,pc\,cm$^{-3}$ would exceed Planck's optical depth constraints \citep{beniamini}. We do not explicitly forecast the CHR/FRB survey's ability to constrain reionization with FRBs, because it will depend on 
$K$-corrections, the intrinsic redshift distribution, and the compute available to search at $\rm DM>4000$\,pc\,cm$^{-3}$. We do, however, argue that such events are likely to exist and will prove useful if they can be detected.

The MeerTRAP survey discovered FRB\,20240304B with $\rm DM = 2458.20$\,pc\,cm$^{-3}$ at $z=2.148$ and a final S/N of 114.7 \citep{caleb2025fastradioburst3}. It was associated with a star-forming galaxy with $M_*\sim10^7\,M_\odot$.
This $z>2$ source was discovered out of fewer than 20 total  
MeerTRAP's coherent-mode localizations, indicating that high-$z$ FRBs are not rare above that flux density. Scaling from the 
sensitivity of MeerKAT, DSA could have detected 
FRB\,20240304B at $z>9$, assuming a spectral index greater than $-1.5$ for $K$-corrections. 
Similarly, a number of DSA-110 FRBs discovered at $z<1.5$ would be detectable by the DSA at $z>5$, due to its $60\times$ higher point-source sensitivity \citep{law2024, sharma_preferential_2024, connor2025}. It is an open question if the Universe produces FRBs at such early times. 
If FRBs have similar engines to long GRBs, we expect to find a significant sample of HeII and HI-reionization FRBs. 
Conversely, if FRBs arise from progenitor channels with substantial delay times, 
such as older magnetar populations, compact-object interactions, or other evolved stellar systems, 
then the volumetric FRB rate may decline rapidly at high redshift. Current evidence appears to contradict this scenario. 
Given the preferential occurrence of FRBs 
in star-forming galaxies \citep{james2022, sharma_preferential_2024, caleb2025fastradioburst3}, 
we expect sources beyond He-reionization.

\subsubsection{Cosmology \& fundamental physics}\label{sec:cosmology}

\vspace{3mm}
\noindent \textit{Large-scale structure:} 
FRB DMs are unusual amongst cosmic probes because a large fraction of the measurement is signal. For a typical source at $z=1$ and DM$=1100$\,pc\,cm$^{-3}$, ``noise'' from the Milky Way and FRB host galaxy 
ISM will contribute less than 15$\%$ to the total observed DM. 
The remaining DM contains signal because the IGM and intervening halos trace the large-scale structure, and even host halo gas is sensitive to cosmology and feedback. 
In contrast, an individual weak lensing galaxy's shear measurement is 99$\%$ shape noise and a pixel of an SZ map contains little information on its own. \citet{Shion_2026} have argued 
that FRB DMs are an unbiased tracer of the large-scale 
structure for $k\lesssim 0.1\,h\,\rm Mpc^{-1}$; even in 
strong feedback scenarios with significant baryon suppression, baryons trace the total matter distribution at linear scales. 
Combining these arguments, a relatively small number of FRB sightlines could have similar constraining 
power on the linear matter power-spectrum as probes with many more 
sightlines \citep{ebina26}. \citet{Shion_2026} show that $10^5$ 
FRBs have roughly equivalent statistical power to $10^8$ 
weak-lensing galaxy shape measurements at large scales. 
Cross correlation of DSA FRBs with galaxies and other tracers will enable constraints on 
$\sigma_8$ and $\Omega_b$. Furthermore, a sample of 
$\sim$\,10$^5$ may be sufficient to detect the 
baryon acoustic oscillations (BAO) with FRB DMs if S/N of 50–100 
can be cumulatively reached across cross correlation bins 
near $r_{\perp}\sim100\,h^{-1}\,\rm Mpc$.
An accurate forecast would require modeling the redshift distribution and 
total residual DM variance, which we leave to future work.

\citet{Madhavacheril_2019} proposed breaking the optical depth degeneracy inherent to kSZ measurements with FRB DMs, allowing for the full exploitation of kSZ tomography. The signal measured by kSZ is $\frac{\Delta T_{\rm CMB}}{T_{\rm CMB}} \propto \int n_e \, v_{\parallel} \, dl$, where $v_{\parallel}$ is the bulk flow of electrons along the line of sight, dominated by the cosmological velocity field. The authors show that a direct measurement of optical depth with FRBs will constrain the cosmic growth rate, $f\sigma_8$, which is sensitive to massive neutrinos, dark energy perturbations, and modifications of General Relativity \citep{Madhavacheril_2019}. A joint kSZ/FRB measurement has recently been done for the first time by \cite{mccarty2026b} with a sample that is 0.2$\%$ of the expected DSA sample.

\vspace{3mm}

\noindent \textit{Strong gravitational lensing:} In addition to measuring diffuse, baryonic matter, 
FRBs provide a window into compact dark matter via strong lensing \citep{munoz16, Eichler17, connorravilens}. 
Lensed FRBs will show up as multiple copies of the same burst. Only $\sim$\,10$^{-3}$ sources will be strongly lensed, necessitating many unique sightlines. 
FRBs are a unique strong lensing case because of their short duration and coherent nature \citep{Eichler17, Wucknitz21}. These facts allow for extraordinary precision in measuring lensing time delays, and the possibility to search for small lensing objects 
with sub-millisecond delays. By saving phase-preserving voltage data for each FRB, multiple copies of the same pulse can be discovered via autocorrelation down to the inverse bandwidth, $1/B$ \citep{leung_lensing, kader_2022}. A positive signal at non-zero time lag 
distinguishes a lensed FRB from different pulses emitted by a repeating source.

For lensing time delays shorter than a typical DSA pointing, deflector masses 
could be detected with $5\times10^{-2}\,M_\odot \lesssim M_{lens} \lesssim 10^{8}\,M_\odot$, assuming a minimum of 1\,$\mu$s in the offline lensing search. This would produce leading extragalactic constraints on primordial black holes \citep{greenpbh}. CHR/FRB will search for compact dark matter sub-halos, MACHOs, and intermediate mass black holes. 
Individual stars at cosmological distances could produce hundreds of $\sim50$\,$\mu$s strong lensing events in the DSA sample \citep{connorravilens}. 
For longer timescales and higher halo masses, it may be possible to model the deflector mass sufficiently 
well to predict arrival times of lens copies. A lensed FRB should also have a strongly lensed host galaxy, 
and detailed O/IR follow up could be used to build a mass model. Such events can be used 
to constrain $H(z)$ via time-delay cosmography \citep{Wucknitz21}. 

\vspace{3mm}

\noindent \textit{Photon mass:} A nonzero photon rest mass $m_\gamma$ would produce an energy-dependent group velocity and hence a delay across frequencies  \citep{wu2016,shao17,bonetti2017},
\begin{equation}
\Delta t_\gamma =
\frac{m_\gamma^2 c^4}{2 h^2 H_0}
\left(
\nu_{\rm low}^{-2} - \nu_{\rm high}^{-2}
\right)
H_\gamma(z),
\qquad
H_\gamma(z) =
\int_0^z
\frac{dz'}{(1+z')^2 E(z')},
\end{equation}
where $E(z)=H(z)/H_0$. Because this delay has the same $\nu^{-2}$ dependence as cold-plasma dispersion, the two contributions are degenerate for an individual FRB. They can nevertheless be separated statistically through their distinct redshift kernels. The mean intergalactic dispersion measure scales as
\begin{equation}
{\rm DM}_{\rm cos}(z)
\propto
\int_0^z
\frac{(1+z')f_e(z')}{E(z')},dz',
\end{equation}
whereas the photon-mass kernel scales as $(1+z)^{-2}E(z)^{-1}$ and is nearly saturated by $z \simeq 2$. A large sample of host-associated FRBs spanning a broad redshift range can therefore break the degeneracy. Alternatively, a conservative approach can be adopted where total time-delay is used as an upper limit on the photon-mass induced dispersion. The redshift distribution of DSA FRBs benefits photon mass constraints. The high-$z$ tail is
disproportionately valuable because beyond $z \approx 2$ the photon-mass
kernel is nearly flat in redshift while $\mathrm{DM}_{\rm IGM}$ continues to grow, making the most distant DSA bursts the 
most valuable for $m_\gamma$.

\vspace{3mm}

\noindent \textit{Targets for a cosmic positioning system (CPS):} The CPS experiment is a recently-proposed solar system
scale VLBI array whose goal is to detect wavefront curvature from FRBs and geometrically measure distances to sources at cosmological distances without the distance ladder \citep{booneCPS,mcquinn2026niacprojectreportsolar}. The experiment requires high-frequency, repeating FRBs whose scattering screens are not resolved by the 40-AU baselines. The DSA's large sample and sub-arcsecond localization precision could provide a list of targets for such a mission. The wide-band DSA FRB survey will tightly constrain spectral index, repetition rate, and scattering/scintillation properties, and may identify globular cluster FRBs without contamination from the host ISM.

\vspace{3mm}

\subsection{The physical origin of FRBs}
The chief advantages of DSA for 
understanding FRB origins are the sample size within a large cosmic volume, wide redshift distribution, localization precision, and simultaneous deep radio imaging. These properties enable 
population studies along multiple axes (repetition, polarization, host galaxies, environments, etc.) and the discovery of rare, informative sources (e.g. periodic repeaters, multiwavelength transients, unusual environments) that can be studied in detail. The DSA is not optimal for the discovery of the nearest, brightest FRB sources, as this requires maximizing FoV. Such sources may include those best suited for the identification of prompt multiwavelength counterparts \citep{Chen2020}. On the other hand, the DSA is well suited for the discovery of low-luminosity source classes that may be close to or below the thresholds of existing instruments \citep[e.g.,][]{KirstenM81}, including events from predicted or known locations / times such as the prompt counterparts to neutron-star mergers and neutron-star / black-hole mergers (Kosogorov et al., in prep.), and to other cataclysms such as giant magnetar flares \citep{Tendulkar2016} and white-dwarf mergers \citep{Shariat2026} among many others. 

\subsubsection{Host galaxy properties}
FRBs have been discovered in a variety of host galaxy types, 
spanning several dex in stellar mass, metallicity, and galactic offsets \citep{law2024, bhandari21, sharma_preferential_2024, gordon25}. While  
the heterogeneity distinguishes them from short GRBs, long GRBs, and 
superluminous supernovae, the FRB host galaxy distribution is not uniform. \citet{sharma_preferential_2024} showed a preference for star-forming hosts, with a deficit of low-mass 
galaxies relative to core-collapse supernovae and total star formation. There is one 
confirmed globular cluster host \citep{bhardwajM81, KirstenM81} and one massive and quiescent elliptical host galaxy \citep{Eftekhari_queiscent, shah25}. The latter has a large offset (40\,kpc) and 
shows no evidence of local scattering or significant host DM. Such outliers evoke speculation about multiple FRB progenitor channels.

FRB host galaxy studies are currently limited by sample size and localization precision. The CHR/FRB will address both issues, increasing the galaxy sample by orders of magnitude and improving localization precision by at least $5\times$ over the existing sample's median. Population studies will answer the offset distribution question and measure stellar masses and metallicities for thousands of FRB host galaxies. Extrapolating from the current sample, we expect to detect sub-populations with $\gtrsim100$ sources in globular clusters, massive elliptical hosts, and dwarf galaxies. We will learn what preferences exist 
between FRB radio properties (host DM, polarization, repetition, luminosity, etc.) and host galaxy types.

\vspace{3mm}
\subsubsection{FRB Repetition}
FRBs fall into two observational classes: repeaters and apparent non-repeaters. It is an open question if repeaters
are fundamentally distinct from once-off FRBs or if all 
FRBs repeat with a broad range in repetition. However, there 
is now strong evidence that radio pulse morphology 
differs between the two classes. Repeating FRBs are 
wider in time and narrower in frequency; most non-repeaters 
are broadband and relatively short duration \citep{Pleunis_2021b, frbcollaboration2026secondchimefrbcatalogfast}. Short periodicity has not been observed despite periodicity searches down to 1\,ms on multiple active sources \citep{spitler14, hessels19, Du_2024}. Periodic activity, however, is established in at least one source (FRB\,20180916A), which is highly active for several days every 16.3\,days and quiescent otherwise \citep{R3,Pastor_Marazuela_2021,Pleunis_2021}. 

The sensitivity and broad frequency coverage of CHR/FRB 
will dig deeply into the per-FRB luminosity function and help determine spectrotemporal and polarimetric 
properties of repeating sources. Repeating sources from CHIME/FRB have a shallower DM-implied redshift distribution than non-repeaters \citep{The_CHIME_FRB_Collaboration_2023_repeaters, chime_rn4}, caused in part by the selection effect of only being able to detect the high-$L$ tail of the luminosity function for a fixed flux limit \citep{wenbin2020}. The CHR/FRB survey will therefore find high-$z$ repeaters relative to the existing sample, answering questions about their redshift evolution and preferred host galaxies 
over cosmic time. 

Fast periodicity searching on the DSA will 
be enabled by the pulsar timing backend, which will record 
high time- and frequency-resolution data for a single beam 
placed on a repeating source. Most sky positions will have 
an observing cadence of days to years, limiting 
trials periods to shorter than a pointing, or days and longer. Possible deep drilling and pulsar timing fields, however, will receive more exposure with much higher cadence. Such fields are more likely to produce new periodically active FRBs and 
tight constraints or detections of fast periodicity. The absence of periodicity associated with neutron star rotation is itself interesting, suggestive of a `Christmas tree'-like emission scenario without a co-rotating radio beam at fixed longitude.

\vspace{3mm}

\subsubsection{Persistent radio sources}

The first FRB to be localized to its host galaxy was the repeating source, FRB\,20121102A, which was found to be coincident ($\lesssim0.1''$) with a 
compact ($\lesssim0.7$\,pc), continuum radio source \citep[$0.4-30$\,GHz;][]{spitler14,chatterjee_direct_2017,tendulkar17}. Such persistent radio sources (PRSs) 
have been found at the sites of a small, but informative, subset of FRBs \citep{Niu2022,moroianu_milliarcsecond_2025}. Thus far, PRSs have tended to appear alongside repeating FRBs in star forming dwarf galaxies, often with high host RM and DM \citep{michilli18}. Key open questions remain about the nature of 
FRB PRSs. It is unknown if they are nebulae directly associated with a neutron star, such as a magnetar wind nebula, a supernova remnant, or if they are a compact H\textsc{ii} region near which the FRB-emitting object was born. It is also unclear if they are related to ``wandering black holes'' or low-luminosity AGN \citep{Eftekhari_2020}. It is not known why they have been detected in dwarf galaxies with repeating FRBs. 

The simultaneous deep, high dynamic range imaging and FRB discovery on the DSA will help answer a number of these questions. Compact continuum objects that are coincident with CHR/FRB sources 
will be studied above $\sim$\,2\,$\mu$Jy at $0.7-2$\,GHz. 
The stacked DSA radio camera images could detect 
FRB\,20121102A's PRS at $z\approx2$, though identifying it as compact on VLBI scales would be difficult.
Such objects should also be detected independently of FRBs, as some will not host FRB-emitting objects or their FRBs will be beamed away from us. The implied volumetric rate of GHz compact continuum sources is 
large \citep{lawPRS} and candidates discovered by the DSA could be followed up with VLBI. Without VLBI, 
PSF modeling, scintillation studies, and super-resolution imaging \citep{polish} will place 
upper limits on the PRS size to distinguish them from compact star formation \citep{chenPRS}. Additionally, the RMs of PRSs observed with the 
DSA radio camera will be compared against FRB RMs. With Euclidean sources counts, we expect $10^4$ PRS-like objects to be discovered by the DSA radio camera at $z<0.5$; the number of FRB/PRS pairs remains to be seen. 
\vspace{3mm}

\subsubsection{Extragalactic pulsars \& magnetars}
There is no formal definition of an FRB. The de facto category includes short-duration ($<1$\,s) radio pulses from beyond our galaxy
\citep[e.g.][]{2016MNRAS.457..232C}. 
Even if they are physically distinct, there 
is likely a continuum between giant pulses from Galactic neutron stars and cosmological FRBs, but the luminosity function's slope between 
$L_\nu = 10^{25}$\,erg\,s$^{-1}$\,Hz$^{-1}$ (a typical Crab giant pulse) 
and $L_\nu = 10^{34}$\,erg\,s$^{-1}$\,Hz$^{-1}$ (a $z=1$ FRB) is an open question. The FRB-like event from Galactic magnetar SGR\,1935+2154 was $L_\nu = 10^{29}$\,erg\,s$^{-1}$\,Hz$^{-1}$ \citep{Bochenek2020STARE2, Andersen2020SGR1935CHIME}, bridging the gap between known Galactic phenomena and FRBs. The globular cluster source, FRB\,20200120E, is 
also lower luminosity than the typical $z\ge0.10$ FRB 
by a few orders of magnitude \citep{KirstenM81}.

The DSA could detect FRB\,20240428A from SGR\,1935+2154 at $d\lesssim100$\,Mpc, given the factor of $\sim10^5$ in sensitivity compared to STARE2 \citep{Bochenek2020STARE2}. CHR/FRB therefore has the opportunity to detect giant radio magnetar bursts in the local universe. CHR/FRB could also detect giant pulses from extragalactic pulsars in nearby galaxies like M31 and M81, filling in the luminosity gap from 
pulsars to cosmological FRBs. 

\vspace{3mm}

\subsubsection{FRB polarization}
FRB emission is coherent and often highly polarized. FRBs exhibit a wide range of phenomena in their spectrotemporal polarization properties. This includes bursts that are nearly 100$\%$ linearly polarized \citep{shermanpol}, bursts that have no detectable polarization, Stokes V detections \citep{shermanpol}, pulsar-like position angle (PA) swings \citep{liu2025polarizationpositionangleswing}, flat intra-pulse PAs \citep{michilli18}, an anti-correlation between pol fraction and frequency in repeaters \citep{feng_depol}, as well as Faraday conversion and standard Faraday rotation \citep{michilli18, HilmarssonRM, cho_pol, shermanRM}. While the microphysics of these 
observations is not understood, the short timescales and similarities with pulsar emission point towards a magnetospheric origin. 

CHR/FRB will preserve polarization information by storing voltage data at the raw $\sim8\,\mu$s and $\sim122$\,kHz resolution. This sample will precisely measure the distribution of RMs, polarization fractions, PA swings, and generalized Faraday conversion. The observations can be tested against models of the FRB emission mechanism. Polarization measurements can also be correlated against host galaxy and local environment properties, thanks to the sub-arcsecond localization precision.

\vspace{3mm}

\subsubsection{Probing plasma turbulence and source sizes with scattering \& scintillation}

FRBs ubiquitously experience multipath propagation from refraction and scattering (diffraction) caused by  small-scale ($\sim 100$\,km to tens of au) electron density fluctuations along the LOS. Scattering manifests as angular broadening in the image domain, pulse broadening in the time domain, and scintillation in both the time and frequency domains. These observables are highly sensitive to the power spectrum of electron density fluctuations, including its spectral index, amplitude, and characteristic scales. While angular broadening of extragalactic sources is typically $\ll100$ mas at 1 GHz \citep[e.g.][]{koryukova2022}, and hence will not be detectable in imaging except for FRB sightlines through particularly dense regions of the Galaxy, both pulse broadening and scintillation will be routinely observable with DSA. Measuring these two effects in tandem is especially powerful, as it enables localization of scattering screens along the LOS \citep{masui2015}. 

While FRB scintillation appears to primarily arise from the Milky Way \citep[e.g.][]{sammons2023}, the bulk of FRB pulse broadening times that are observed require scattering contributions from extragalactic plasma, likely within host galaxies and the circumsource medium \citep{chawla2022,ocker2022}. However, the exact origin of FRB scattering and its relation to other burst and sightline properties, including polarization/RMs, host galaxies, and potentially other intervening galaxies along the LOS, remain open questions. The pulse broadening time  scales with frequency approximately as $\propto \nu^{-4}$, and hence will increase by more than a factor of 10 between the highest and lowest frequency bands of the FRB search; the scintillation bandwidth decreases by the same factor.
This range underlies the ability to  distinguish scattering from the intrinsic  frequency dependence of emitted bursts.   The DSA FRB survey will thus be sensitive to pulse broadening times and scintillation bandwidths spanning multiple orders of magnitude, enabling a wide range of science cases: cross-correlating scattering with host galaxy properties (including redshift) \citep{glowacki2025}, probing any scattering from the tenuous, multiphase circumgalactic medium \citep{faber2024,jow2026}, and using scattering to probe dense circumsource nebulae and emission regions \citep{ocker2023, 2024MNRAS.527..457K, nimmo2025}. Scattering at high redshifts ($z>2$) will be a particularly exciting frontier for DSA, as the nature of turbulence and multi-phase gas in the IGM (as well as ISM and CGM) may substantially differ from  that at $z<1$ \citep{lapiner2026}. 

DSA’s high detection rates will also reveal rare phenomena like plasma lensing events, including burst echoes and extreme (de)-magnification \citep{cordes2017,kader2026}. Combining these detections with the commensal imaging survey, which will yield thousands of plasma lensing events towards background AGN, will resolve the elusive astrophysics of these tiny-scale plasma structures and inform searches for even more elusive gravitational lensing (Section~\ref{sec:cosmology}). 

\section{Conclusions}
\label{sec:concl}
We have presented the Chronoscope FRB survey on the Deep Synoptic Array, focusing on forecasts and science applications. The DSA's 1650$\times$6.15\,m antennas and low $T_{sys}$ will have similar point-source sensitivity to FAST, but with $>$\,100$\times$ the FoV. 
Three sub-bands will each form and search several million beams at millisecond timescales up to a maximum DM of $\sim$\,5000\dmunits. This amounts to 
the largest-ever FRB search by a wide margin. We have predicted detection rates in each sub-band, 
the redshift distribution of observed events, and 
the magnitudes of their host galaxies. We have highlighted several key science cases. Below, we offer a list of central takeaways.

\begin{itemize}
    \item The CHR/FRB survey is expected to detect roughly $10^4$ bursts per year in each sub-band. The total number of detections across sub-bands could reach $10^5$ in the full five-year survey, dependent on source counts, spectral index, and available compute. Reducing available compute to $10\%$ of the baseline would decrease detection rates by just a factor of a few. 
    \item Thanks to the essentially filled aperture of the DSA, the well-characterized PSF, and simultaneous reference imaging from the Radio Camera, most CHR/FRB events will be localized at $\lesssim250$\,mas. Events with S/N\,$\gtrsim100$ are expected to occur multiple times per week, enabling precision of several tens of milliarcseconds.
    \item The DSA sensitivity will result in a deep redshift distribution, with a median of $z$ $\sim$\,1, dependent on cosmic evolution of the population. We expect a significant sample at $z>2$. FRBs from the epoch of reionization will be easily detectable if progenitor delay times are comparable to the age of the Universe at $z>5$.
    \item Host galaxy follow up will require a broad range of public and targeted surveys, including both photometric and spectroscopic redshifts, spanning from radio to optical. Host galaxy identification and characterization of CHR/FRB sources will benefit greatly from DESI, Argus Array, Rubin/LSST, Roman, Euclid, Via/Boombox, JWST, ALMA, and large ground-based facilities. The DSA/HI survey will provide kinematics and redshifts for many FRB hosts at $z\lesssim0.30$. Most DSA FRBs will not have spectroscopic redshifts, but a significant fraction will have photometric redshifts.
    \item Nearly all FRB science cases are accelerated by a large sample of bursts localized with sub-arcsecond precision. The application of FRBs to baryonic feedback and cosmology will benefit the most, as each additional sightline contributes information about cosmic gas, magnetic fields, and the large-scale structure. Questions surrounding FRB origins will be addressed by large samples of sources and host galaxies extending to high redshifts, and the discovery of rare FRBs that can be studied in detail.
\end{itemize}

\section*{Acknowledgments}

This research is enabled in part by development funded by Schmidt Sciences for the Deep Synoptic Array project (DSA). The DSA is part of the Eric and Wendy Schmidt Observatory System in partnership with the California Institute of Technology. 

\bibliographystyle{apsrev4-1}

\bibliography{oja_template}




\end{document}